\documentclass[]{aa}  

\usepackage{graphicx}
\graphicspath{{Plots}}

\usepackage{txfonts}
\usepackage{float}
\usepackage{placeins}

\usepackage{natbib,twoopt}
\usepackage[breaklinks=true,draft]{hyperref} 
\bibpunct{(}{)}{;}{a}{}{,}             

\newcommand{\kms} {km\,s$^{-1}$}
\newcommand{\kmspixel} {km\,s$^{-1}$ pixel$^{-1}$}

\newcommand{\Msun} {$\mbox{M}_{\sun}$}
\newcommand{\Mstar} {$\mbox{M}_{\ast}$}

\newcommand{\ppix}{\,pixel$^{-1}$}
\newcommand{\id} {HDFS-ID\,\#}
\newcommand{\idu} {UDF10-ID\,\#}
\newcommand{\degree} {$^\circ$\,}
\newcommand{\RN}[1]{\textup{\uppercase\expandafter{\romannumeral #1}}}

\newcommand{\atlas}{{ATLAS$^{\rm 3D}$}}
\newcommand{\lr}{{$\lambda_R$}}
\newcommand{\manga}{MaNGA }

\newcommand{\udft}{\textsf{udf-10}}
\newcommand{\cprep}{Contini~et~al.,~in~prep.}
\newcommand{\cprepp}{Contini~et~al.~(in~prep.)}

\newcommand{\Inami}{I17-paper II}
\newcommand{\Inamip}{(I17-paper II)}

\newcommand{\Bacon}{B17-paper I}
\newcommand{\Baconp}{(B17-paper I)}

\defcitealias{Contini2016}{C16}

\begin{document} 

\title{The MUSE Hubble Ultra Deep Field Survey:\\ \RN{5}\,. Spatially resolved stellar kinematics of galaxies at redshift $0.2\lesssim z \lesssim 0.8$  \thanks{Based on observations made with ESO telescopes at the La Silla-Paranal Observatory under programmes 094.A-0289(B), 095.A-0010(A), 096.A-0045(A) and 096.A-0045(B).}}
\titlerunning{The MUSE HUDFS V: Resolved stellar kinematics of intermediate redshift galaxies}

\author{Adrien~Gu\'{e}rou
          \inst{1,2,3}
          \and
          Davor~Krajnovi\'{c}
          \inst{4,\thanks{\email{dkrajnovic@aip.de}}}
          \and
          Benoit~Epinat
          \inst{1,2,5}
          \and
          Thierry~Contini
          \inst{1,2}
          \and
          Eric~Emsellem
          \inst{3,6}
          \and
          Nicolas~Bouch\'{e}
          \inst{1,2}
          \and
          Roland~Bacon
          \inst{6}
          \and
          Leo Michel-Dansac
          \inst{6}
          \and
          Johan~Richard
          \inst{6}
          \and
          Peter~M.~Weilbacher
          \inst{4}
          \and
          Joop Schaye
          \inst{7}
          \and
          Raffaella Anna Marino
          \inst{8}
          \and
          Mark den Brok
          \inst{8}
          \and
          Santiago Erroz-Ferrer
          \inst{8}
          }

\institute{
                        IRAP, Institut de Recherche en Astrophysique et Plan\'{e}tologie, CNRS, 14, avenue Edouard Belin, F-31400 Toulouse, France
                \and
                        Universit\'{e} de Toulouse, UPS-OMP, Toulouse, France
        \and
                        European Southern Observatory, Karl-Schwarzschild-Str. 2, D-85748 Garching, Germany
                \and
                Leibniz-Institut f\"{u}r Astrophysik Potsdam (AIP), An der Sternwarte 16, D-14482 Potsdam, Germany
        \and
                        Aix Marseille Universit\'{e}, CNRS, LAM, Laboratoire d’Astrophysique de Marseille, UMR\,7326, 13388 Marseille, France
                \and
            Univ Lyon, Univ Lyon1, ENS de Lyon, CNRS, Centre de Recherche Astrophysique de Lyon UMR5574, F-69230, Saint-Genis-Laval, France
        \and
                Leiden Observatory, Leiden University, P.O. Box 9513, 2300 RA Leiden, The Netherlands
        \and
            ETH Zurich, Institute of Astronomy, Wolfgang-Pauli-Str. 27, CH-8093 Zurich, Switzerland
}

   \date{Received 31 March 2017 ; accepted 26 June 2017}

  \abstract
  {We present spatially resolved stellar kinematic maps, for the first time, for a sample of 17 intermediate redshift galaxies ($0.2\lesssim z \lesssim 0.8$). We used deep MUSE/VLT integral field spectroscopic observations in the Hubble Deep Field South (HDFS) and Hubble Ultra Deep Field (HUDF), resulting from $\approx$\,30\,h integration time per field, each covering 1\arcmin\,$\times$\,1\arcmin\, field of view, with $\approx$\,0\farcs65 spatial resolution. We selected all galaxies brighter than 25\,mag in the I band and for which the stellar continuum is detected over an area that is at least two times larger than the spatial resolution. The resulting sample contains mostly late-type disk, main-sequence star-forming galaxies with $10^{\,8.5}$\,\Msun\,$\lesssim$\,\Mstar\,$\lesssim 10^{\,10.5}$\,\Msun. Using a full-spectrum fitting technique, we derive two-dimensional maps of the stellar and gas kinematics, including the radial velocity $V$ and velocity dispersion $\sigma$. We find that most galaxies in the sample are consistent with having rotating stellar disks with roughly constant velocity dispersions and that the second order velocity moments $V_{\rm rms}=\sqrt{V^2+\sigma^2}$ of the gas and stars, a scaling proxy for the galaxy gravitational potential, compare well to each other. These spatially resolved observations of the stellar kinematics of intermediate redshift galaxies suggest that the {\it regular} stellar kinematics of disk galaxies that is observed in the local Universe was already in place 4\,--\,7\,Gyr ago and that their gas kinematics traces the gravitational potential of the galaxy, thus is not dominated by shocks and turbulent motions. Finally, we build dynamical axisymmetric Jeans models constrained by the derived stellar kinematics for two specific galaxies and derive their dynamical masses. These are in good agreement (within 25\,\%) with those derived from simple exponential disk models based on the gas kinematics. The obtained mass-to-light ratios hint towards dark matter dominated systems within a few effective radii.}
  
  \keywords{Galaxies: formation - Galaxies: evolution - Galaxies: Kinematics and dynamics - Galaxies: Stellar content}
  \maketitle
  
\section{Introduction}
\label{sec:introduction}
The kinematics of galaxies is of paramount importance for our understanding of their formation and evolution. Large three-dimensional spectroscopic surveys such as GHASP~\citep{Epinat2008a,Epinat2008b}, \atlas~\citep{Cappellari2011a}, SAMI~\citep{Croom2012,Bryant2015}, CALIFA~\citep{Sanchez2012, Garcia-Benito2015}, the ongoing \manga survey~\citep{Bundy2015}, and the MUSE Atlas of Disks (MAD; Carollo et al.,~in prep.) have intensively characterised, or will intensively characterise, the stellar and ionised gas kinematics of local galaxies ($z\approx$\,0) over a large range of galaxy masses ($\sim$\,$10^{\,8.5\,--\,11}$\,\Msun). These surveys have shed light on the importance of mergers~\citep{Arnold2014, Naab2014, Haines2015} and gas accretion~\citep{Davis2011, Cheung2016} in the growth of galaxies that we see today. The build-up of stellar disks has also been studied through the specific stellar angular momentum of early-type galaxies~\citep[e.g. with proxies such as \lr,][]{Emsellem2011FRSR}. 

Similar integral field spectroscopic (IFS) surveys have been performed over the past decade to target high-redshift galaxies ($z$\,$\approx$\,1\,--\,3). Still limited by the performance of state-of-the-art instruments, such as for example\, SINS~\citep{Cresci2009}, MASSIV~\citep{Contini2012}, LSD/AMAZE~\citep{Gnerucci2011}, KMOS$^{3D}$~\citep{Wisnioski2015}, and KROSS~\citep{Stott2016}, the kinematics of such galaxies have been characterised mostly (if not only) by their ionised gas content. At these redshifts, most galaxies are gas rich and a large portion of their gas is transformed into stars, which has led to the known peak of the cosmic star formation rate (SFR)~\citep[$z$\,$\approx$\,1\,--\,2.5;][]{Hopkins2006, Madau2014}. This gas could be brought to galaxies through, for example\, cold gas accretion at $z\geq 2$~\citep{Keres2005, Dekel2009, VandeVoort2011} or major and minor mergers~\citep{Lin2008, DeRavel2009, Lopez-Sanjuan2011, Lopez-Sanjuan2013} with a suggested peak of merging events between $1\leq z\leq 2$~\citep{Ryan2008, Conselice2008}. However, merger events might not be the dominant channel~\citep{Kauffmann2010}. 

The observed gas kinematics of high-redshift galaxies are often perturbed~\citep{Epinat2012camel} by,  for example clumpy star formation sites~\citep{ForsterSchreiber2009, Swinbank2012} and strong stellar feedback~\citep{Dib2006, Green2010}. As a result, the ionised gas velocity dispersion in the disk component of high-redshift galaxies is about 5 to 10 times larger than in the local Universe ($\geq$\,60\,\kms, as compared to $\approx$\,10\,--\,20\,\kms; e.g.\, \citealt{Epinat2010, Erroz-Ferrer2015}) and the origin of the gas velocity dispersion is still under debate. A constant evolution (decrease) of the ionised gas velocity dispersion is observed from $z$\,$\approx$\,2 to $z\approx$\,1 and seems to follow the gas fraction of galaxies~\citep{Wisnioski2015}, as expected from disk stability theory~\citep{Toomre1964}.

In order to understand the physical processes that transform such clumpy disk star-forming galaxies at high redshift into the ordered galaxies of the local Universe, recent studies have been conducted on intermediate redshift galaxies ($0.2\leq z \leq 0.7$); these studies include~DEEP2~\citep[][with multislit spectroscopy]{Kassin2007,Miller2014}, IMAGES~\citep[][with GIRAFFE/VLT IFS]{Puech2010}, MUSE-HDFS~\citep[][with MUSE/VLT IFS]{Contini2016}, and very recently extended up to $z\approx$\,1.7 with MUSE and KMOS~\citep{Swinbank2017}. Such galaxies have been found to be mostly rotation dominated~\citep[][from their ionised gas]{Contini2016, Swinbank2017} and a clear evolution towards a settlement of disk components has been suggested from $z\approx$\,1 to today~\citep{Kassin2012}. Similar ionised gas kinematics have been observed at all probed stellar masses ($\sim$\,$10^{\,8-10}$\,\Msun) and the Tully-Fisher relation, linking the maximal rotation velocity to the luminosity (mass) of a galaxy, is found to extend over the full range of galaxy stellar mass~\citep{Miller2011, Contini2016}. However, a larger scatter is observed at lower stellar masses, which is consistent with the larger fraction of low-mass galaxies that are not yet ``settled'' at $z\approx$\,0.2\,--\,1, in contrast with most massive galaxies~\citep{Kassin2012, Simons2015}. 

Important steps in the understanding of the kinematic evolution of galaxies have been made through these numerous intermediate- and high-redshift galaxy surveys. However, gas (the main tracer used to infer galaxy kinematics at these redshifts) is a complex ingredient that leads to significant uncertainties in the observed kinematics, such as turbulence, heating and cooling, clumps, and inflows or outflows. The stellar content of galaxies is thus a more robust tracer of the gravitational potential as it overcomes some of these issues. Resolved spectroscopy of the stellar continuum of intermediate-redshift galaxies ($0.2 \leq z \leq 1$) is a challenging task and has so far mostly been accessible with long-slit spectroscopy, for example for single galaxies~\citep{VanderMarel2007, VanderWel2008} or stacks of large samples of galaxies~\citep{Shetty2014}. Such studies have led to the characterisation of the star formation history (SFH) of intermediate-redshift galaxies ($z\approx$\,0.8), which was found to be similar to local galaxies, and to the first estimate of the dark matter (DM) content of the most massive galaxies ($\sim$\,10$^{\,11}$\,\Msun), which was found to be low within 1.5 R$_e$~\citep{Shetty2014}. In the more distant Universe, long-slit spectroscopy has revealed the stellar content of a handful of very massive quiescent galaxies (\Mstar\,$\geq 10^{\,11}$\,\Msun) at $z\approx$\,1~\citep{Belli2014a} and $z\approx$\,2~\citep{VanDokkum2009, VandeSande2011, Toft2012, VandeSande2013, Belli2014a}. Such studies have shown that the stellar velocity dispersion is generally higher than in the local Universe, i.e.\, $\approx$\,300\,\kms~, and could be as high as $\approx$\,500\,\kms~for some specific galaxy~\citep{VanDokkum2009}. The size\,--\,mass diagram of distant galaxies has also been intensively investigated using their dynamical mass that can be estimated from the measurement of their velocity dispersion. Various studies have led to the finding that galaxies at $z$\,$\approx$\,1\,--\,2, at fixed mass, are more compact than galaxies in the local Universe~\citep{Daddi2005, Trujillo2006, VanDokkum2008, Cappellari2009, VanDokkum2010}.

Except for a few gravitationally lensed galaxies at redshift $z$\,$\approx$\,2~\citep{Newman2015,2017ApJ...838...14M}, IFS observations of the stellar content of galaxies further away than the local Universe was missing until now. With the state-of-the-art IFS MUSE instrument~\citep{Bacon2010}, a new window has now been opened on the stellar kinematics of intermediate-redshift galaxies. Indeed, the large field of view (FoV) of MUSE and its high sensitivity allow one to perform deep blind observations. Such an observation strategy leads to new detections of distant galaxies~\citep{Bacon2015} and to a significant increase of the exposure time on large samples, favouring the stellar continuum detection of distant objects.\\ 

This paper is organised as follows. In Sect.~\ref{sec:chap4-data_sample} we present the deep MUSE observations performed in two Hubble Deep fields along with the sample selection and global properties. In Sect.~\ref{sec:kinematics}, we describe the kinematics analysis and we present our results in Sect.~\ref{sec:results}, focusing on the comparison between the stellar and gas kinematics. In Sect.~\ref{sec:dicussion} we discuss our results and their implications for the commonly used assumptions on the kinematics of such intermediate-redshift galaxies. Finally, Sect.~\ref{sec:summary} summarises our work.

\section{Data sets and galaxy sample}
\label{sec:chap4-data_sample}

We used the two deepest data sets of MUSE observations for the present study: one targets the Hubble Deep Field South (HDFS), presented in~\cite{Bacon2015}; and the other, the Ultra Deep Field-10 (\udft), presented in \citet[][hereafter \Bacon]{Bacon2017}, is located in the Hubble Ultra Deep Field (HUDF; \citealt{Beckwith2006}). A summary of the data reduction processes and the resulting properties of the data cubes are presented in the following paragraphs.   

\subsection{MUSE observations in the HDFS}
\label{subsec:MUSE_obs_hdfs}

\begin{figure*}
        \begin{center}
        \includegraphics[width=\columnwidth]{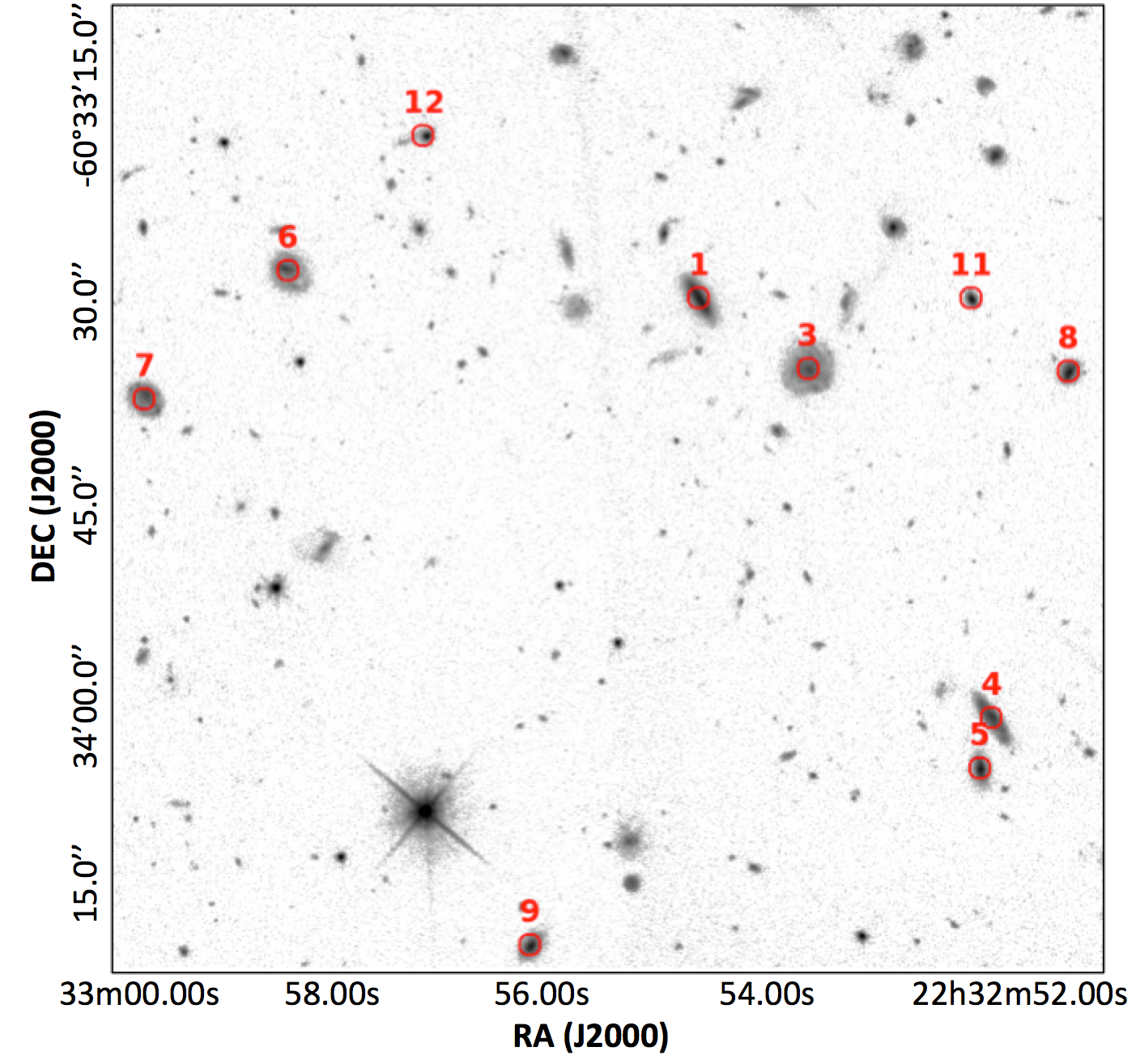}
    \includegraphics[width=\columnwidth]{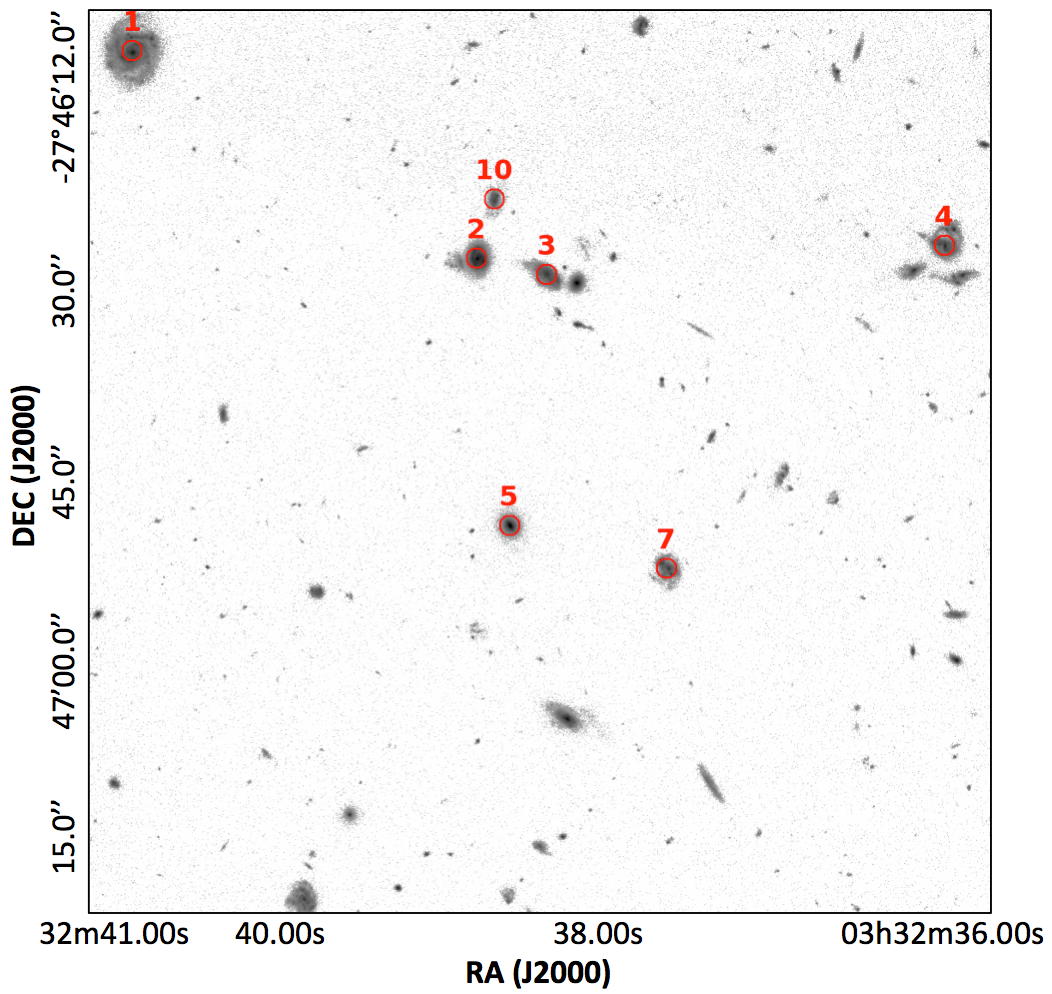}
        \caption{$-$ HST/WFC2 F814W image of the Hubble Deep Field South (HDFS, left panel), and the Hubble Ultra Deep Field-10 (\udft, right panel) with the locations of the galaxies comprising our sample of spatially resolved continuum galaxies. The sizes of the circles correspond to the MUSE spatial resolution ($\approx0.66$\arcsec~for the HDFS, and $\approx0.63$\arcsec\, for the \udft). The identification numbers are those from the catalogues of \cite{Bacon2015} and \Inami, also listed in Table~\ref{tab:gal_prop}.}
        \label{fig:deep_fields}
    \end{center}
\end{figure*}

The MUSE observations in the HDFS were obtained during the last commissioning\footnote{ESO programme 60.A-9100\,(C)} of the instrument on the unit 4 telescope (UT4) at the Very Large Telescope (VLT). These observations were presented in \cite{Bacon2015} and consist of a total exposure of 27 hours, covering a field of 1\arcmin\,$\times$\,1\arcmin\, and reaching a 1$\sigma$ emission-line surface brightness limit of 1\,$\times$\,10$^{-19}$ erg\,s$^{-1}$\,cm$^{-2}$\,arcsec$^{-2}$. The HST image of the HDFS region observed with MUSE is presented in Fig.~\ref{fig:deep_fields}~(left panel). The redshifts of 189 objects were measured down to an apparent magnitude of $I_{814}=29.5$\,mag, increasing the number of known spectroscopic redshifts by more than an order of magnitude. The redshift distribution spans a wide range, from $z$\,$\approx$\,0\,--\,7, and shows peaks at $z$\,$\approx$\,0.6 and $z$\,$\approx$\,3. The effective spatial resolution of the combined data cube is 0.66\arcsec~at 7000\,\AA\, and is about 10\% better (worse) at the red (blue) end of the MUSE spectral range going from 4750\,--\,9300\,\AA\,. The spatial sampling (i.e.\, spaxel size) is $\approx$\,0.2\arcsec~and the theoretical spectral resolution is $\approx$\,2.3\,\AA\, full width half maximum (FWHM). The observing strategy and data reduction are described in \cite{Bacon2015} and summarised in \cite{Contini2016}, hereafter~\citetalias{Contini2016}. 

The  MUSE HDFS data set is publicly available\,\footnote{\url{http://muse-vlt.eu/science/hdfs-v1-0}} and consists of the fully reduced data cube and the extracted sub-cubes of each of the objects identified in \cite{Bacon2015}. The object masks and variance cubes (derived through the MUSE pipeline) were also made available to the public\,\footnote{\url{http://data.muse-vlt.eu/HDFS/Web/}}. We here used the first public data release of the MUSE HDFS (i.e.\, v1.0).

\subsection{MUSE observations in the UDF-10}
\label{subsec:MUSE_obs_udf}

The MUSE observations of the \udft~in the HUDF (see Fig.~\ref{fig:deep_fields}, right panel), cover an area of 1.15 arcmin$^2$ at a depth of $\approx$31 hours of exposure time and were acquired during Guaranteed Time Observations between September 2014 and December 2015. The final data cube is of much better quality than the HDFS thanks to an improved observational strategy and data reduction scheme; see \Bacon~for the detailed data reduction processes and quality assessment. The effective spatial resolution of the final data cube ranges from 0\farcs57 in the red (at 9350\,\AA) to 0\farcs71 in the blue (at 4750\,\AA). The $1\sigma$ emission-line surface brightness limit is 2.8\,$\times$\,10$^{-20}$ ergs s$^{-1}$ arcsec$^{-2}$ in the red (6500$-$8500\,\AA) between OH sky lines.

The redshift of 252 objects were securely measured and span the range of $0.21\leq z \leq 6.64$ \citep[][hereafter \Inami]{Inami2017}. The 50\% completeness is reached at 26.5~mag (F775W) and 32 detected objects do not have prior counterpart in the HST catalogue of~\cite{Rafelski2015}. This increases the number of spectroscopic redshifts in this area of the HUDF by
almost one order of magnitude.

\subsection{Galaxy sample: Selection criteria}
\label{subsec:sample_selection}
We searched in the MUSE catalogues of~\citet[][HDFS]{Bacon2015} and~\Inami~(\udft) for galaxies that are spatially resolved and have a bright enough continuum so that the main absorption lines (i.e.\, Balmer lines, CaK, Mgb, and Fe) have a signal-to-noise ratio (S/N) that is suitable for extracting reliable resolved stellar kinematics from full spectral fitting. 

To identify such sample, we first selected the galaxies brighter than I$_{814W/850LP}=25$\,mag. This corresponds to the magnitude limit above which the Balmer absorption lines are no longer detected in the spatially integrated galaxy spectrum. From this sample, we then selected galaxies that have a stellar continuum with a S/N of at least 1 per spectral pixel in each spectrum over an area of 16 MUSE spaxels, i.e.\, that corresponds to about 1.5 times the PSF FWHM size of the data cube. The S/N ratios were estimated between 4150\,--\,4350\,\AA\, (in the object rest frame), representative of the continuum level (i.e. almost free of absorption and emission lines) based on the variance cubes produced by the MUSE pipeline. For galaxies that ended up in the final sample, the S/N ratio of the central 16 spaxels was typically between 5 (e.g. for galaxy \id10) to 10 (e.g. for galaxy \id4) per pixel. A subsample of 30 objects (15 in the HDFS and 15 in the \udft) was thus obtained. To ensure a robust extraction of the stellar kinematics, we then excluded 13 galaxies (5 in the HDFS, 8 in the \udft) for which we could not obtain a spatially binned data cube with more than six bins with a S/N greater than 8 (see \S\ref{subsec:binning}). This criterion results in a subsample of 17 galaxies (10 in the HDFS and 7 in the \udft) with a spatially resolved continuum in the two MUSE data cubes. This sample is presented in Table~\ref{tab:gal_prop}. 

The 10 galaxies selected in the HDFS MUSE data cube were all included in the sample of~\citetalias{Contini2016}, who studied their spatially resolved gas kinematics based on the same data set. We use their analyses as a reference for comparison with our stellar and gaseous kinematics. We also use the HDFS catalogue of~\citet{Bacon2015}, which comprises~\citetalias{Contini2016} selected objects whose strongest emission line (i.e.\, [OII], [OIII], or H$\alpha$) covers an area larger than 20 spaxels (i.e.\, twice the seeing disk) with a S/N per pixel above 15. These authors thus obtained a sample of 28 galaxies. No prior selection on magnitude was applied, as opposed to the sample defined here, but a redshift cut at \textit{z}\,$\approx$\,1.5 was imposed because of the loss of the ``strong'' emission lines in the MUSE wavelength range above this redshift. Applying the same criteria to the \udft~ has led to a sample of 36 spatially resolved galaxies suitable for analysis of the gas kinematics (\cprep). For comparison, this sample is also shown in Figures\,\ref{fig:mass_redshift} and \ref{fig:mass_sfr}.

\subsection{Galaxy sample: Global properties}
\label{subsec:global_prop}
Here we describe the redshift, stellar mass, and SFR distributions of the galaxy sample and the morphologies and close environment derived from HST images. The redshifts come from \cite{Bacon2015} and \Inami~catalogues for the the HDFS and \udft~fields, respectively. The stellar masses and SFRs were estimated using the stellar population synthesis (SPS) code FAST~\citep{Kriek2009}, as described in~\citetalias{Contini2016}. For \udft~we used the extended UV-to-NIR HST photometry of \cite{Rafelski2015}.

\begin{figure}
        \includegraphics[width=\columnwidth]{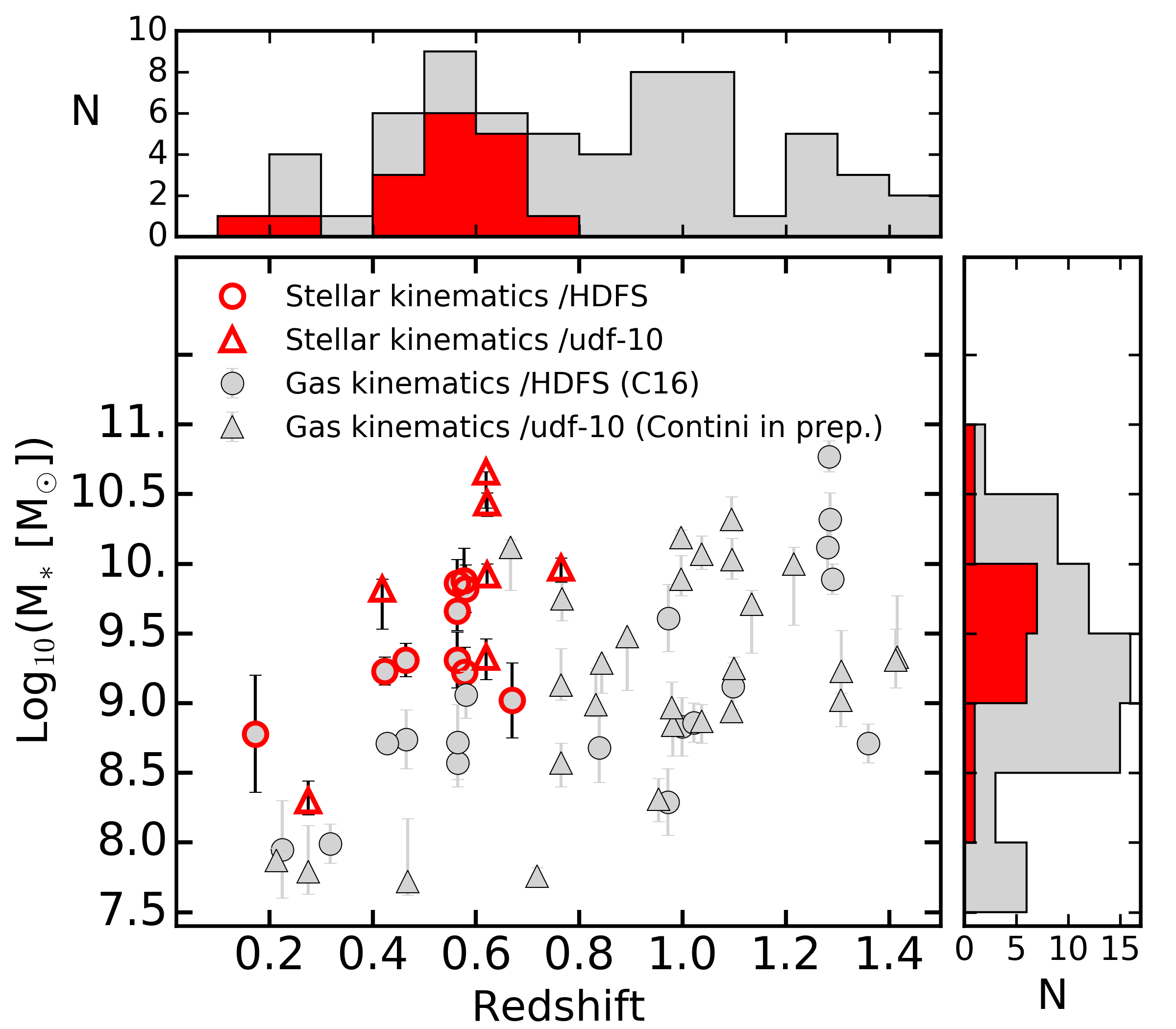}
        \caption{Distribution of stellar mass as a function of redshift for the sample of spatially resolved continuum galaxies (red markers). Galaxies suitable for the analysis of spatially resolved gas kinematics (\citetalias{Contini2016}; \cprep) are indicated by the grey symbols. Above redshift $z=$\,0.8, no galaxies are spatially resolved in their continuum owing to the natural decrease of the apparent magnitude with redshift and significant OH sky line contamination in the MUSE wavelength range at these redshifts.}
        \label{fig:mass_redshift}
\end{figure}

Figure\,\ref{fig:mass_redshift} shows that the galaxy sample is spread over the redshift range 0.17\,$\lesssim z \lesssim$\,0.76. The lack of galaxies at redshift $z\gtrsim$\,0.8 is due to the decrease of the apparent magnitude with redshift but also (and mainly) a consequence of significant OH sky line residuals in the MUSE data cubes that degrade the S/N in the continuum spectra, even for the most massive galaxies at $z$\,$\approx$\,1.2 (see Fig.\,\ref{fig:mass_redshift}). The majority (11/17) of the sample galaxies are located at a redshift $z$\,$\approx$\,0.6, which reflects the observed peaks in the redshift distributions of the MUSE HDFS~\citep{Bacon2015} and \udft~\Inamip~fields. 

In terms of stellar mass, the galaxy sample ranges between $\sim$\,$10^{\,8.5}$\,\Msun~and $10^{\,10.7}$\,\Msun, where the majority of the sample (13/17) have a stellar mass between $\sim$\,$10^{\,9}$\,\Msun~and $10^{\,10}$\,\Msun. At a given redshift, the stellar continuum can only be spatially resolved with sufficient S/N ratios for the most massive objects among the sample of galaxies suitable for gas kinematics (Contini et al. 2016; \cprep). Nevertheless, as in \citetalias{Contini2016}, this sample probes the low-mass regime of the intermediate-redshift galaxy population studied so far with IFS, such as IMAGES~\citep[$z$\,$\approx$\,0.4\,--\,0.75;][]{Puech2010} or KMOS-HIZELS~\citep[$z$\,$\approx$\,0.8;][]{Sobral2013} in a similar redshift range. 

\begin{figure}
        \includegraphics[width=\columnwidth]{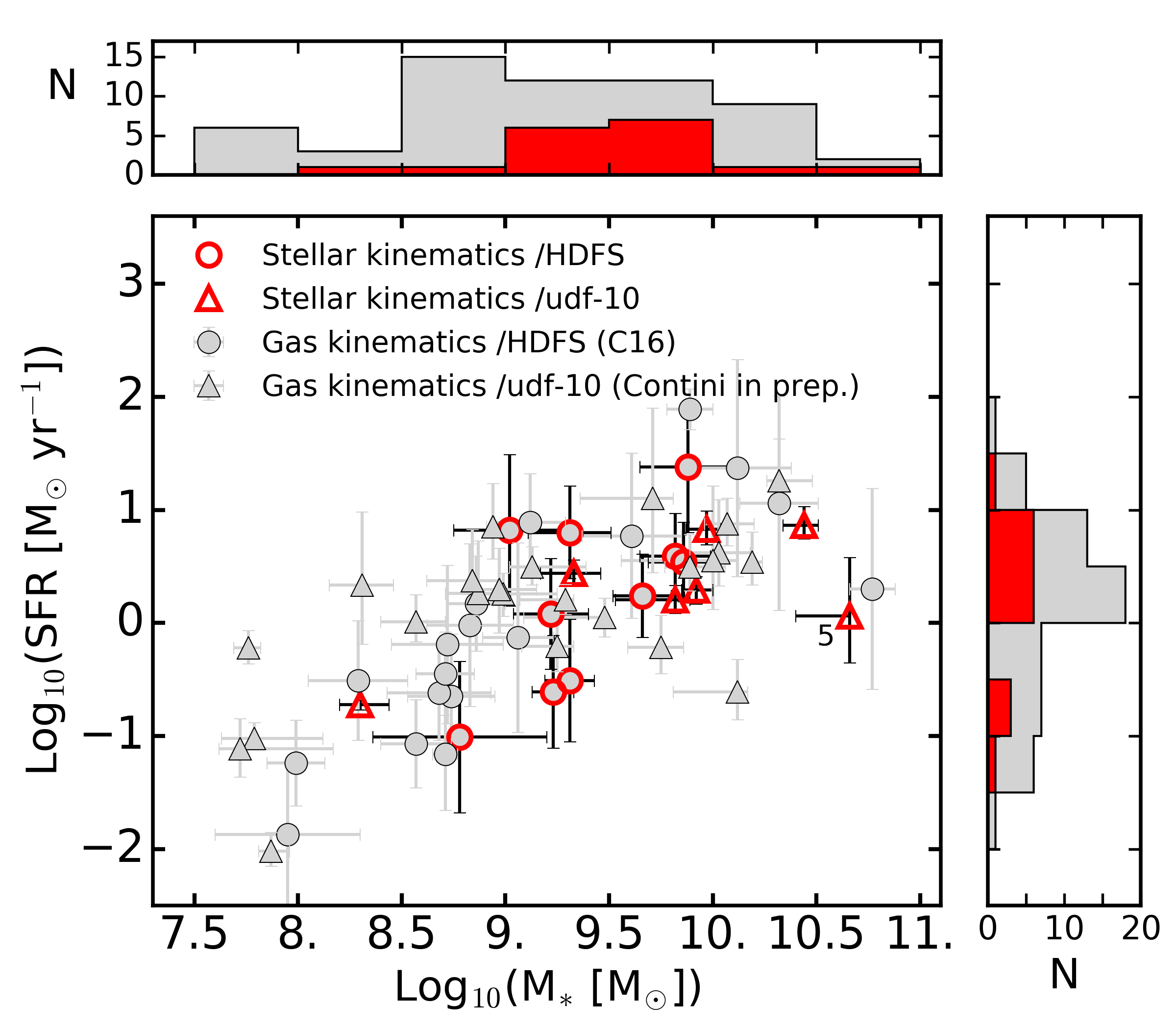}
        \caption{Distribution of the SFR as a function of the stellar mass for the sample of spatially resolved continuum galaxies (red markers). Galaxies suitable for the analysis of spatially resolved gas kinematics  (\citetalias{Contini2016}; \cprep) are indicated by the grey symbols. The sample galaxies fall along the ``normal'' star-forming sequence of galaxies and extend to the lowest stellar mass regime ($\sim$\,$10^{\,8.5}$\,\Msun) probed so far with IFS surveys. \idu5, indicated by its ID number, shows no ionised gas in the MUSE data cube (except a weak [OII] line) and deviates from the ``normal'' star-forming sequence.}
        \label{fig:mass_sfr}
\end{figure}

\begin{table*}
\caption{Galaxy sample global properties}
\label{tab:gal_prop}
\centering
\begin{tabular}{crccclclll} 
\hline\hline\\
Field & ID & \textit{z} & F814W / F850LP & R$_{e}$ & \textit{i} & log$_{10}$(\Mstar) & log$_{10}$(SFR) & Morphology - Notes\\
- & - & - & (mag) & (arcsec) & ($^\circ$) & ([\Msun]) & ([\Msun\,$yr^{-1}$]) & -\\ 
(1) & (2) & (3) & (4) & (5) & (6) & (7) & (8) & (9)\\
\hline\\
HDFS & 1 & 0.17 & 21.22 & 0.87 & 75.0 & 8.78$\pm$0.42 & -1.01$\pm$0.67 & -\\
HDFS & 3 & 0.56 & 21.52 & 1.34 & 16.0 & 9.66$\pm$0.14 & 0.24$\pm$0.37  & Spirals - clumpy\\
HDFS & 4 & 0.56 & 21.78 & 1.38 & 75.0 & 9.86$\pm$0.17 & 0.54$\pm$0.35  & Weak spirals\\
HDFS & 5 & 0.58 & 21.97 & 0.37 & 68.0 & 9.82$\pm$0.17 & 0.59$\pm$0.38  & Compact - Nucleus - weak spirals\\
HDFS & 6 & 0.42 & 21.98 & 0.60 & 29.0 & 9.23$\pm$0.10 & -0.61$\pm$0.50    & Spirals - bar\\
HDFS & 7 & 0.46 & 21.99 & 0.73 & 41.0 & 9.31$\pm$0.12 & -0.51$\pm$0.54 & Clumps - Spirals?\\
HDFS & 8 & 0.58 & 22.08 & 0.30 & 61.0 & 9.88$\pm$0.23 & 1.38$\pm$0.58   & Satellite\\
HDFS & 9 & 0.56 & 22.08 & 0.42 & 61.0 & 9.31$\pm$0.20 & 0.8$\pm$0.41    & Compact\\
HDFS & 11 & 0.58 & 22.72 & 0.17 & 62.0 & 9.22$\pm$0.18 & 0.08$\pm$0.49 & Compact\\
\medskip
HDFS & 12 & 0.67 & 22.79 & 0.09 & 37.0 & 9.02$\pm$0.27 & 0.82$\pm$0.67  & Compact - ``jet''\\
UDF-10 & 1 & 0.62 & 20.13 & 1.26 & 27.0 & 10.44$_{-0.1}^{+0.07}$ & 0.86$_{-0.12}^{+0.16}$ & Grand design spiral\\
UDF-10 & 2 & 0.42 & 20.72 & 0.54 & 34.0 & 9.82$_{-0.29}^{+0.07}$ & 0.21$_{-0.12}^{+0.13}$ & Asym. spiral - bar - merger?\\
UDF-10 & 3 & 0.62 & 21.42 & 0.75 & 61.0 & 9.92$_{-0.07}^{+0.08}$ & 0.29$_{-0.13}^{+0.12}$ & Asym. spiral - merger?\\
UDF-10 & 4 & 0.76 & 21.48 & 0.72 & 50.0 & 9.97$_{-0.1}^{+0.07}$ & 0.83$_{-0.14}^{+0.16}$  & Asym. spiral - Clump - merger?\\
UDF-10 & 5 & 0.62 & 21.25 & 0.36 & 49.0 & 10.66$_{-0.26}^{+0.0}$ & 0.06$_{-0.42}^{+0.52}$ & Early-type disk-like\\
UDF-10 & 7 & 0.62 & 21.91 & 0.66 & 50.0 & 9.33$_{-0.16}^{+0.13}$ & 0.44$_{-0.11}^{+0.12}$ & Asym. spiral - bar\\
UDF-10 & 10 & 0.27 & 22.49 & 0.45 & 57.0 & 8.3$_{-0.1}^{+0.14}$ & -0.72$_{-0.05}^{+0.04}$ & Stellar stream - satellite\\
\hline
\end{tabular}
\tablefoot{Properties of the sample of galaxies with spatially resolved continuum\,: (1)~MUSE data set (2)-(3)~Galaxy ID number, and redshift, from~\citet{Bacon2015} and \Inami~catalogues~(3)-(4)~Apparent magnitude I$_{814W}$ (HDFS) and I$_{850LP}$ (\udft) (5)~Effective radii from \citet[][HDFS]{Casertano2000} and \Inami~(\udft) (6)~Galaxy inclination from~\citetalias{Contini2016}~(HDFS) and \Inami~(\udft) (7)-(8)~Stellar mass and SFR measured in~\citetalias{Contini2016}~(HDFS) and \Inami~(\udft) (9)~Morphological properties from HST images.}
\end{table*}

Figure\,\ref{fig:mass_sfr} shows the SFR of the galaxy sample as a function of the stellar mass. The SFRs span over two orders of magnitudes from SFR\,$\approx$\,0.1\,\Msun\,yr$^{-1}$ to $\approx$\,25\,\Msun\,yr$^{-1}$, which is almost the entire range of SFR covered by the sample of \citetalias{Contini2016}, except for the few most extreme cases. As pointed out in \citetalias{Contini2016} and confirmed in \cprep, deep MUSE observations in the HDFS and \udft~fields allow us to probe a new class of objects at these intermediate redshifts, both in terms of stellar mass and SFR, i.e.\, lower mass (\Mstar\,$\leq$\,$10^{\,9.5}$\,\Msun) and fewer star-forming galaxies (SFR\,$\leq$\,10\,\Msun\,yr$^{-1}$) than in previous IFS surveys. As shown in~\citetalias{Contini2016}, the empirical relation between the SFR and stellar mass for ``normal'' star-forming galaxies (defined in~\citealt{Whitaker2014} from the CANDELS fields) can be broadly extended to lower mass and SFR regimes (see their Fig.\,3). The sample of galaxies defined here is therefore part of the ``normal'' star-forming sequence of galaxies at these intermediate redshitfs. One galaxy, \idu5, stands out of this sequence with a low SFR (log$_{10}$(SFR [\,\Msun\,yr$^{-1}$])$\approx0.06$) for a stellar mass of $\sim$10\,$^{10.5}$\,\Msun\,, for example\, it has no ionised gas detected in the MUSE data cube, except for a weak [OII] emission line.

We used the deep HST images obtained in the HDFS and \udft~fields to assess the morphology and close environment of our galaxy sample. The morphological types are dominated by late-type disk galaxies (14 of 17). Eight of these galaxies have clear spiral arms visible in high-resolution HST images (see Table~\ref{tab:gal_prop}) and are either disturbed, asymmetric, or exhibit a bar. Two other disk galaxies appear very inclined (i.e.\, i\,$\approx$\,75$^\circ$; \id1 and \id4) and elongated, whereas three others are rather compact (i.e.\, $\lesssim$\,1.5\arcsec\, in diameter). The last late-type disk galaxy, \id8, potentially has a satellite in the north-east side, clearly visible in its HST image~\citepalias[see also][their Appendix B]{Contini2016}. The last three galaxies of the sample do not show clear ``disk-like'' structures. For example, \id12 is rather small and round with a jet-like structure that extends towards the east; \idu10 exhibits an irregular morphology with numerous large clumps, and a prominent stream extending towards the north; and finally, \idu5 is a clear early-type disk-like galaxy (see Appendix.~\ref{fig:kin_maps5_udf}) and does not show any ionised gas (except a weak [OII] emission line), suggesting that it is a quiescent field galaxy at redshift $z\approx$\,0.6. We point out that \id9 has also been suggested to be an early-type disk-like galaxy with a prominent bulge~(\citealt{Bacon2015}, \citetalias{Contini2016}). 

Despite the limited number of galaxies in our sample, we witness a great variety in the morphological properties of these intermediate redshift galaxies. The global properties of the sample, such as redshift, stellar mass, SFR, disk inclination, effective radii, and morphological peculiarities, are summarised in Table~\ref{tab:gal_prop}. Disk inclination and effective radius (R$_e$) for the HDFS and \udft~galaxies are extracted from \citetalias{Contini2016} and \citet{VanderWel2014}, respectively.

\section{Stellar and gas kinematics}
\label{sec:kinematics}

\subsection{Characterisation of the MUSE spectral resolution}
\label{subsec:lsf}
A critical ingredient in a robust measurement of the intrinsic stellar kinematics of a galaxy using the full spectral fitting technique, and in particular its velocity dispersion (i.e.\,the broadening of the spectral lines), accounts for the instrument spectral resolution, also called the line spread function (LSF). Indeed, the stellar (and gas) templates used to fit the observed spectra are spectrally convolved to match the instrumental spectral resolution to separate the instrumental broadening of the lines from
the intrinsic velocity dispersion of the observed galaxy.

The MUSE pipeline provides a measure of the LSF that is achieved on single calibration files (e.g.\, arc exposures). The MUSE HDFS and \udft~data cubes consist of a combination of more than 50 single dithered exposures rotated by multiples of 90 degrees. This observational strategy (i.e.\, rotation and dithering) significantly helps build master flats and lower systematics and leads to a spatial homogeneisation of the LSF of the MUSE data cube towards an average of the instrumental resolution. We thus decided to accurately characterise the LSF by relying on a direct measurement on the final combined data cubes, following the same methods as in the first paper of this series~\Baconp.

We used the sky emission lines of the combined non sky-subtracted MUSE data cubes, not corrected for heliocentric velocity, to characterise the instrumental LSF. At first order, we assumed the shape of the LSF to be Gaussian and measured the FWHM of 19 groups of 1\,--\,10 bright sky emission lines spread as uniformly as possible over the MUSE wavelength range. These lines were selected from the UVES catalogue.~\citep{Hanuschik2003} which provides their expected intensities and FWHM. We performed this analysis for each spaxel of the MUSE cubes using CAMEL\footnote{\url{https://bitbucket.org/bepinat/camel.git}}~\citep{Epinat2012camel} and corrected the expected intrinsic FWHM of the sky lines given by the UVES catalogue. We then fitted the relation FWHM\,--\,$\lambda$ with a second order polynomial using a 2$\sigma$ clipping rejection and thus obtained the MUSE LSF for each spaxel of the HDFS and \udft~data cubes. 

\begin{figure}
        \begin{center}
                \includegraphics[width=\columnwidth]{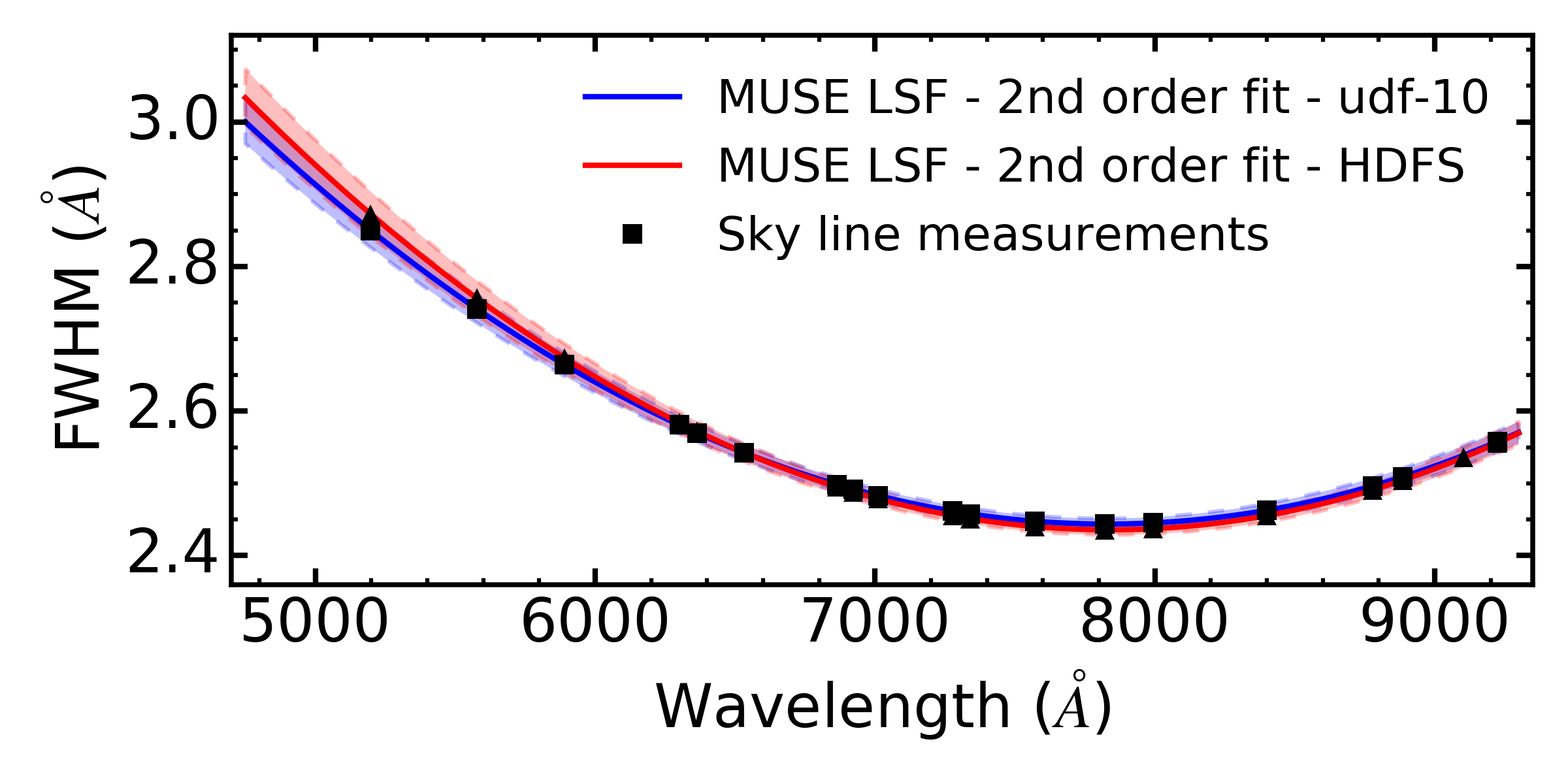}
                \caption{MUSE line spread functions (LSF) measured in the non sky-subtracted HDFS and \udft~MUSE data cubes. The LSF spatial mean FWHM (i.e.\, averaged over the data cube {\bf FoV}) are indicated by the solid lines, and their 1$\sigma$ spatial variation by the respective shaded areas. The latter corresponds to a typical error in the derived velocity dispersion of $\sigma\lesssim$\,1\,\kms. The black squares (\udft) and triangles (HDFS) indicate the positions of the sky line measurements.}
                \label{fig:lsf_measurement}
        \end{center}
\end{figure}

We find a mean spatial variation (i.e.\, from spaxel to spaxel, within a given data cube, for a given sky line) of the LSF FWHM of 0.05\,\AA\,, both for the HDFS and \udft~MUSE data cubes, which corresponds to a typical error in the derived velocity dispersion of $\sigma\lesssim$~1\,\kms~(for the given galaxy sample). We therefore took the respective spatial mean of the MUSE LSFs (i.e.\, averaged over the MUSE FOVs) as the inferred MUSE spectral resolution function for each respective MUSE data set\,\footnote{For the HDF-S: FWHM($\lambda$)=$6.266 \times10^{-8} \lambda^{2} - 9.824 \times 10^{-4} \lambda +6.286$\\For the \udft: FWHM($\lambda$)=$5.866 \times 10^{-8} \lambda^2 - 9.187 \times 10^{-4} \lambda + 6.040$}. The derived instrumental LSFs FWHM are shown in Fig.\,\ref{fig:lsf_measurement} and correspond to a spectral resolution of R\,$\simeq$\,1600\,--\,3600, and $\sigma_{LSF}$\,=\,70\,--\,40\,\kms~from the blue to the red end, consistent with the nominal instrument characteristics\footnote{\url{www.eso.org/sci/facilities/paranal/instruments/muse/inst.html}}. We underline the high stability of MUSE LSF from these two data cubes, observed at different periods, as detailed in \Bacon.

\subsection{Kinematics extraction: Method}
\label{subsec:ppxf}

We used the last version available (V6.0.0) of the penalised pixel-fitting (pPXF) code~\citep{Cappellari2004ppxf, Cappellari2017} to extract the resolved stellar kinematics of the galaxy sample. The pPXF code\,\footnote{\url{www-astro.physics.ox.ac.uk/~mxc/software/\#ppxf}} uses a stellar library to fit the observed spectrum with a combination of stellar templates. One can thus recover the line-of-sight velocity distribution (LOSVD), namely, the radial mean velocity, \textit{V}, the velocity dispersion, \textit{$\sigma$}, and higher order Gauss-Hermite moments (that characterise the deviation of the distribution from a Gaussian profile) of the observed spectrum. In the present case, we limited ourselves to the two first order moments of the LOSVD, namely, \textit{V} and \textit{$\sigma$}, both for the stellar continuum and gas components (absorption and emission lines). We did not derive higher order moments, such as $h$3 and $h$4, because of the limitations induced by the relatively low S/N ratios and spatial resolution (see \S\ref{subsec:binning}).

The pPXF allows us to simultaneously fit the stellar continuum and gas emission lines of a spectrum, i.e. the minimisation criterion is applied jointly on the stellar and gas fits, but their respective kinematics can be kept independent. Two different sets of templates were thus used to fit the stellar continuum and gas emission lines. We fit the stellar continuum of the galaxies with a subset of 53 templates from the empirical Indo--US stellar library \citep{Valdes2004}. We chose this library because of its spectral resolution of 1.35\,\AA~FWHM~\citep{Beifiori2011}, constant over its full wavelength coverage, i.e. 3460\,--\,9464\,\AA\,, and most importantly, significantly better than the MUSE LSF FWHM (see Section~\S\ref{subsec:lsf}), even for galaxies at $z\approx$\,0.8 once at rest frame. The wavelength coverage is well suited for galaxies at redshift \textit{z}\,$\lesssim$\,0.7 observed with MUSE, as it includes strong absorption lines such as the Balmer series, CaK\,$\lambda$3940, Mgb\,$\lambda$5200, Fe\,$\lambda$5270, and Fe\,$\lambda$5335. The subset of 53 templates has been selected as in~\cite{Shetty2015a} in order to be gap-free and to well represent the library's atmospheric parameters range (T{\rm eff} versus [Fe/H]). Before the fitting procedure, the stellar templates are convolved to the respective MUSE LSF resolution (as defined in \S\,\ref{subsec:lsf}; i.e.\, varying with wavelength). As for the gas components, we fitted the following series of emission lines: H${\eta}\,\lambda$3835, H${\zeta}\,\lambda$3889, H${\epsilon}\,\lambda$3970, H${\delta}\,\lambda$4102, H${\gamma}\,\lambda$4340, H${\beta}\,\lambda$4862, and [OIII]$\,\lambda\lambda$4959,5007. During the fitting procedure, each of these lines is fitted with a Gaussian that is also previously broadened at the respective MUSE LSF resolution (see \S\ref{subsec:lsf}). We did not fit the [OII] doublet emission line as it requires us to include a relatively large wavelength range with very low S/N continuum, which significantly degrades the stellar fit.

\begin{figure*}
        \includegraphics[width=\textwidth]{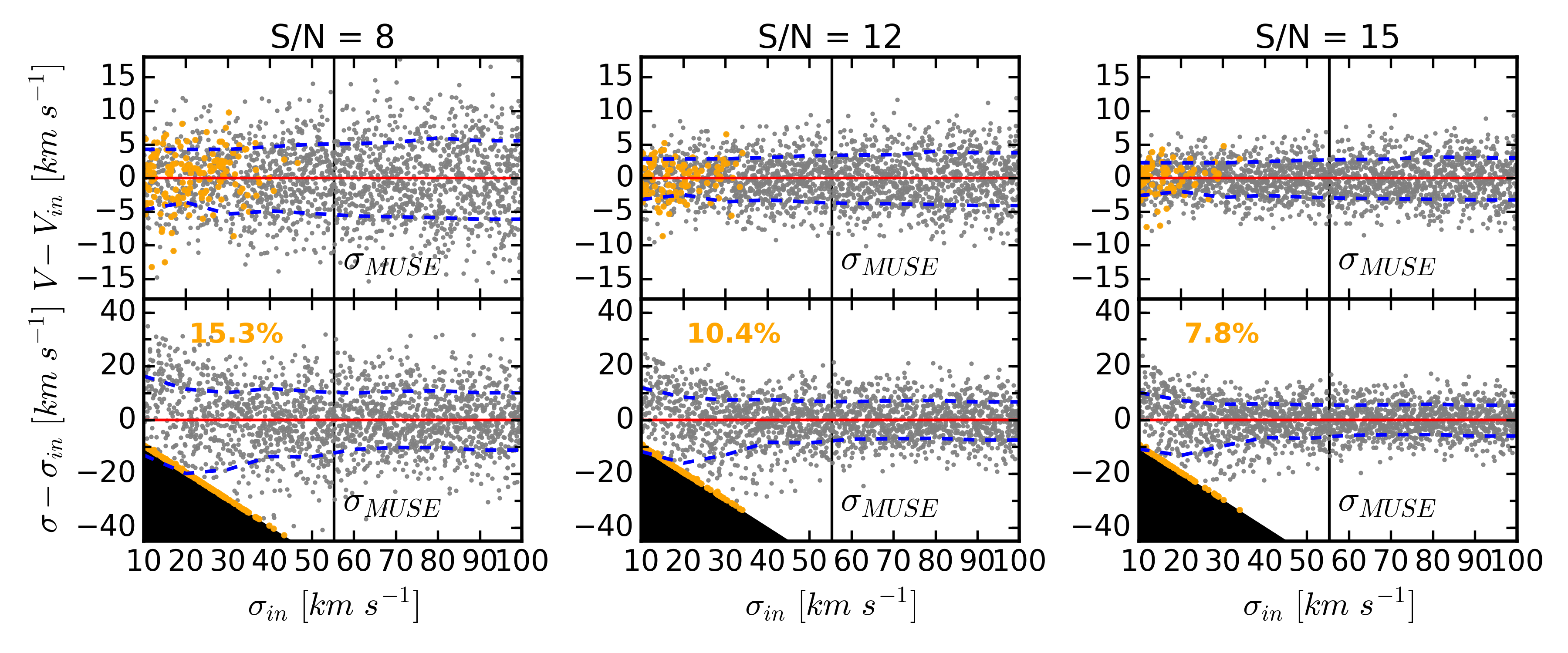}
        \caption{Monte Carlo simulations of the fitting procedure of the MUSE spectra for three different S/N ratios in the stellar continuum, i.e.\, S/N\,=\,8 (left panels), S/N\,=\,12 (middle panels), and S/N\,=\,15 (right panels), representative of the targeted S/N of the spatially binned MUSE observations. We used a stellar template from the Indo-US library~\citep{Valdes2004} that we matched to the MUSE spectral characteristics (i.e.\, wavelength range, spectral resolution, and spectral sampling), and fitted with pPXF using the same procedure as used for the MUSE observations. The top panels show the recovered radial velocity $V$, and the bottom panels the measured velocity dispersion $\sigma$, as a function of the input velocity dispersion $\sigma_{in}$ of the stellar template. The grey points represent 2000 realisations, the blue dashed lines the 1$\sigma$ deviation from the mean value in bins of 10\,\kms, and the black areas represent non-physical solutions. No systematic biases are observed for $V$ at any $\sigma_{in}$ level or for $\sigma$ above $\sigma_{in}\approx$\,40\,\kms. Catastrophic failure events (orange dots) are observed for about 7\,--\,15\% of the cases below $\sigma_{in}\approx$\,40\,\kms. Systematic errors of $\approx$\,5\,\kms~for $V$, and $\approx$\,10\,\kms~for $\sigma$ are observed.}
        \label{fig:simu_sn}
\end{figure*}

Before fitting the MUSE spectra, we spectrally rebinned the spectra in logarithmic scale with a step of $\approx$\,55~\kmspixel, corresponding to the MUSE spectral sampling in velocity space. The stellar templates and gas emission lines were also spectrally rebinned but with a step that is two times smaller, of $\approx$\,27\,\kmspixel, to preserve the highest spectral resolution possible. We fitted the MUSE spectra over the wavelength range 3740\,--\,5100\,\AA\,(in the rest frame of the object), which includes the above-mentioned emission and absorption lines, except for the lowest redshift galaxy of the sample, \id1 ($z\approx$\,0.17), which we could only fit from 4050\,\AA\, but up to 5350\,\AA\, (to include the Mg$b$ line). We masked five strong sky emission lines at 5577\AA, 5889\AA, 6157\AA, 6300\AA, and 6363\AA,\, which potentially contaminate the spectra of the sample galaxies. We allowed different kinematics for the two gas components, i.e. the Balmer series and [OIII] doublet, and imposed a line ratio between [OIII]$\lambda$5007$/$[OIII]$\lambda$4959 of 2.98 as predicted by theory~\citep{Osterbrock1989, Galavis1997}. We set up pPXF to use additive polynomials of the 6th order and multiplicative polynomials of the 1st order, well suited to the wavelength range fitted. Finally, using the set of parameters described above, we first determined the best combination of stellar templates for each object by taking the best-fit solution of the galaxy stacked spectrum (summing all spectra in the MUSE sub-cube belonging to the galaxy), and used this best template to fit each individual spatially binned spaxel (see \S\ref{subsec:binning}). In this way, we reduce the scatter of the kinematic solutions, but potentially introduce systematic biases if there are strong local variations in the stellar population. Again, the gas emission lines were fitted simultaneously with the stellar continuum during this procedure.

\subsection{Spatial binning and kinematics uncertainties}
\label{subsec:binning}


After extracting the best template that fits the stacked galaxy spectrum, we spatially binned the reduced MUSE sub-cubes of each of the sample galaxies to increase and homogenise their spectral S/N. We used the adaptive spatial binning software developed by \cite{Cappellari2003Bin} based on Voronoi tessellation. We estimated the original S/N of each individual spectrum by taking the median S/N on the spectral range 4150\,--\,4350\,\AA\,(in the rest frame of the object), which is representative of the continuum level, i.e.\, almost free of absorption and emission lines. We used the variance cubes associated with the MUSE data cubes, produced by the MUSE pipeline, as a measure of the original noise. 
 
The choice of the target S/N is dictated by the fact that one wants to recover the galaxy kinematics with minimum systematic biases while maximizing the number of individual binned spaxels. The most difficult kinematic parameters to recover are naturally the highest order moments of the LOSVD, i.e.\, the velocity dispersion $\sigma$ in the present case. Based on the gas kinematics analysis of \citetalias{Contini2016}, we can expect, at first order, that the stellar velocity dispersion is of the same order, i.e.\, reaching down to 30\,--\,40\,\kms. These values are below the MUSE LSF resolution (i.e.\,~$\sigma_{MUSE}\geq$\,40\,--\,70\,\kms), but can still be recovered given large enough S/N \citep{Cappellari2016}. To estimate the minimum target S/N that we need to adopt to recover such low velocity dispersion values, we performed Monte Carlo simulations using model spectra tuned to match the spectral properties of the MUSE observations (i.e.\, wavelength range, spectral resolution, and pixel size). We used one of the Indo-US templates as the model spectrum (i.e.\, HD\,120136) and broadened it to 2000 different velocity dispersion values, $\sigma_{in}$, uniformly spread between 10 and 100\,\kms. We then ran pPXF with the same settings as described in Section~\S\ref{subsec:ppxf} and analysed the obtained stellar kinematics. This test was performed for three different values of input S/N, i.e. 8, 12, and 15, similar to the typical highest values of the stellar continuum S/N of the MUSE HDFS and \udft~data cubes. The results are shown in Fig.\,\ref{fig:simu_sn}. This test only probes the systematics of the method but does not account for potential errors due to template mismatch or correlated noise. 

We find no systematic biases in the recovery of $V$ when the MUSE data is binned to a minimum S/N of 8\ppix and find an associated systematic error smaller than $\approx$5\,\kms~for any input velocity dispersion $\sigma_{in}$ (see top panels of Fig.\,\ref{fig:simu_sn}). Regarding the velocity dispersion, Fig.\,\ref{fig:simu_sn} shows (see bottom panels) that binning the MUSE data to a S/N of 8\ppix~is enough to recover velocity dispersion down to 40\,\kms, with an error of $\approx$10\,\kms, regardless of the $\sigma_{in}$ value. However, below $\sigma_{in}\approx$\,40\,\kms, we start to observe catastrophic failure events (i.e.\, the measured velocity dispersion is null) for $\approx$\,15\,\%, 10\%, and 8\% of the cases for an input S/N of 8, 12, and 15\,\ppix , respectively (see orange dots in Fig.\,\ref{fig:simu_sn}).

Therefore, as a compromise between spectral quality and spatial resolution, we chose to bin the MUSE spectra of each galaxy to a median S/N of 15\,\ppix or to the maximal S/N that leads to at least six final spatial bins, typically between 10 and 12\,\ppix\,(see Table~\ref{tab:gal_PA}). Before proceeding with the spatial binning, we rejected all spectra with an initial S/N below 1 and obtained, for a given galaxy, spatially binned spectra with a typical S/N scatter of $\approx$\,1\,--\,4, and made sure that no final bins have  S/N lower than 8\,\ppix. The kinematics analysis (see section \S\ref{subsec:ppxf}) has been performed on these spatially binned sub-cubes and the results are presented in the next section, Section~\S\ref{sec:results}. We do not bin the data spectrally as this would lower the sampling of the LSF (native MUSE sampling is 1.25\AA\, per spectral pixel) and increase the uncertainties on the extracted kinematics.

Finally, we estimated the uncertainties of the derived kinematics by performing 500 fits to each of the individual binned spaxels of the sample galaxies to which we previously added random noise. The level of this noise has been constrained to follow a Gaussian distribution with a sigma equal to the standard deviation of the respective residuals of the original fit. We used the same fitting procedure as described in section \S\ref{subsec:ppxf} and found typical variations of $\Delta V_{s}$\,$\approx$\,10\,\kms~and $\Delta\,\sigma_{s}$\,$\approx$\,20\,\kms~for the stellar kinematics, and $\Delta V_{g}$\,$\approx$\,4\,\kms~and $\Delta\,\sigma_g\approx$\,6\,\kms~for the gas kinematics. The values quoted are averaged over the galaxy spaxels (i.e.\, the galaxy spatial area) and typically decrease (increase) by $\approx$\,50\% for the central (outer) spaxel(s). We also note that the four HDFS galaxies that we spatially binned at a lower spectral S/N (\id7, \id8, \id11, and \id12) and \idu10 (S/N of 10) have higher uncertainties for the stellar velocity dispersion, i.e.\, a mean $\Delta\,\sigma_s$\,$\approx$\,30\,\kms.

\section{Results}
\label{sec:results}

\subsection{Gas kinematics and comparison with \cite{Contini2016}}
\label{subsec:gas_kin}

The gas kinematics of the HDFS galaxy sample has previously been derived in \citetalias{Contini2016} with the use of the python code \textit{CAMEL} \citep{Epinat2012camel}. We compared their results with those obtained here. The main analysis differences between \citetalias{Contini2016} and the present work are as follows. First, in \citetalias{Contini2016} only a small wavelength range around the specific group of gas emission lines is fitted, typically 100\,--\,200\,\AA\,. Second, a constant continuum template is used to fit the stellar continuum (as opposed to the use of a library of stellar templates here). Third,~\citetalias{Contini2016} spatially smoothed the MUSE cubes with a two-dimensional Gaussian of 2 pixels FWHM (enough to obtain the desired S/N ratios on the emission line alone), and therefore, have a significantly better spatial resolution than here. Fourth, the gas kinematics has been independently derived for various groups of emission lines, i.e. [OII] doublet, H\,$\alpha$  , and ([OIII] doublet, H\,$\beta$), as opposed to two independent groups in the present work, i.e. the Balmer series and [OIII] doublet. Both methods use Gaussian templates to fit the gas emission lines. 

\begin{figure}
        \includegraphics[width=\columnwidth]{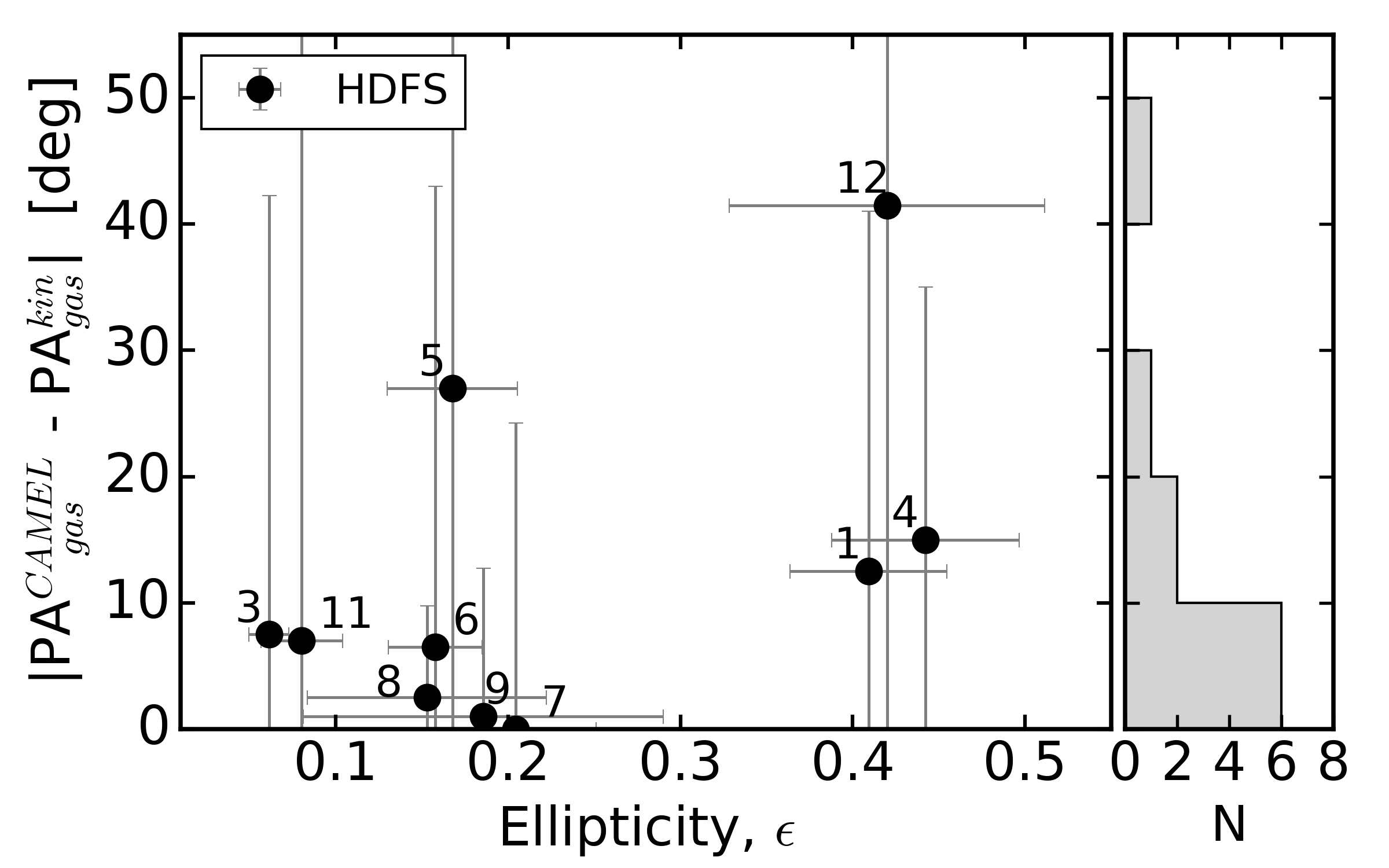}
        \caption{Comparison of the gas kinematic major-axis PA between this work (PA$^{kin}_{gas}$) and \protect{\citetalias{Contini2016}} (PA$^{CAMEL}_{gas}$), as a function of the galaxy's ellipticity measured from the MUSE white light image. Our results are based on the gas kinematics of the brightest emission line (see Table~\ref{tab:gal_PA}) derived with pPXF and the method presented in \protect{\citet{Krajnovic2006}}. Values from \protect{\citetalias{Contini2016}} were measured with CAMEL \protect{\citep{Epinat2012camel}} using the group of lines ([OIII]$\lambda\lambda$5007,4959, H$\beta$). We find good agreement within the (large) uncertainties, with a median $\Delta$PA$_{gas}$ of $\approx$\,7$^\circ$\,$\pm$\,31$^\circ$, and consistent measurements at less than 20$^\circ$ for 8/10 galaxies (see histogram). The galaxy ID numbers are indicated above each point.}
        \label{fig:PA_gas_ppxf_camel}
\end{figure}

We compared the gas kinematic maps presented here, derived from the group of the brightest emission line (see Table.~\ref{tab:gal_PA}) with the results from \citetalias{Contini2016} based on the group ([OIII],~H\,$\beta$)  taking into account the LSF defined here (see \S\,\ref{subsec:lsf}). Despite the obvious lower spatial resolution of the present analysis, we find good agreement between the two methods, i.e. most of the recovered kinematic features (both $V$ and $\sigma$) are similar. We find typical differences of $\Delta$V of a few \kms\ that are consistent within the method uncertainties. However, we observe a systematic positive offset of the gas velocity dispersion values derived with pPXF in comparison to the analysis of \citetalias{Contini2016}. This offset is $\approx$10\,\kms~in the low velocity dispersion regime ($\sigma\leq$\,30\,\kms) and $\approx$\,5\,\kms~for higher velocity dispersion levels ($\sigma\gtrsim$\,40\,--\,50\,\kms). We investigated the origin of this systematic offset and found that it is primarily due to different assumptions made about the continuum level during the fitting procedure. Indeed, we performed a new analysis forcing pPXF to use a constant continuum level (as assumed in \textit{CAMEL}) instead of the library of templates and found better agreement between the two analyses. The remaining differences (less than $\approx$\,5\,\kms~regardless of the velocity dispersion level of the galaxy) might be explained by the difference between the two methods in the spatial binning and the assumptions about the kinematics of the group ([OIII], H$\beta$).    

\begin{figure*}[t]
        \includegraphics[width=\textwidth]{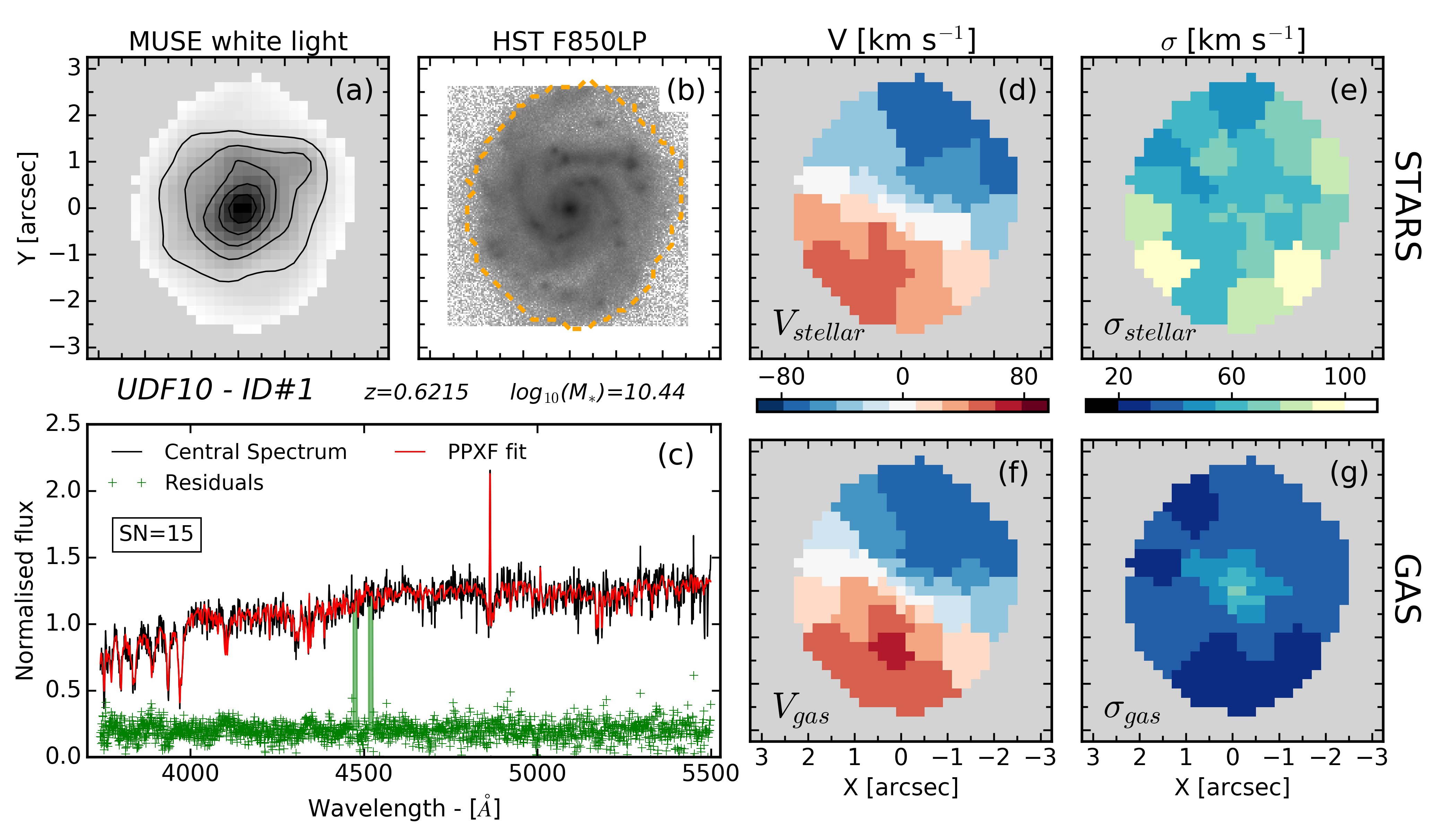}
        \caption{\textit{Kinematics analysis of the \idu1 galaxy.} \textbf{Panel \textit{(a)}} MUSE white light image of the galaxy. Isophote contours are overplotted with black continuous lines. \textbf{Panel \textit{(b)}} HST F850LP image with the spatial coverage of the MUSE data cube (satisfying our spectral S/N rejection criterion) is indicated by the orange dashed contour. \textbf{Panel \textit{(c)}} Fit of the central spectrum of the spatially binned sub-cube (black solid line) and its best fit from pPXF (red continuous line) is shown. The green points represent the fit residuals that are arbitrarily shifted along the vertical axis. The green shaded areas correspond to the masked regions (i.e. sky line residuals).~\textbf{Panels \textit{(d)-(e)}} Resolved stellar kinematics, respectively: the LOS velocity $V_{stellar}$ (panel \textit{d}) and the  velocity dispersion $\sigma_{stellar}$ (panel \textit{e}) are shown. The scales of the colour scheme are indicated by the colour bars at the bottom of each panel. \textbf{Panels \textit{(f)-(g)}} Resolved gas kinematics from the Balmer series of emission lines, respectively: the rotational velocity V$_{gas}$ (panel \textit{f}) and the velocity dispersion $\sigma_{gas}$ (panel \textit{g}) are shown. The scales of the colour scheme are indicated by the colour bars on top of each panel. The galaxy ID number, its redshift, and stellar mass are indicated above panel (c).}
        \label{fig:Main_kin_fig}
\end{figure*}

We also measured the position angle (PA) of the derived gas kinematics, PA$^{kin}_{gas}$, with the code \textit{Fit Kinematic PA}\footnote{\url{www-astro.physics.ox.ac.uk/\~mxc/software/\#pa\_kin}} , which implements the method presented in \cite{Krajnovic2006}. We present the results in Table~\ref{tab:gal_PA} and compare these with the values derived in \citetalias{Contini2016}, based on the {\rm CAMEL} software, PA$^{CAMEL}_{gas}$, in Fig.~\ref{fig:PA_gas_ppxf_camel}, as a function of the ellipticity of the galaxy. The photometric ellipticity has been derived as the first moment of the surface brightness of the MUSE white light image for each object, which was taken at 2\,R$_{e}$ (see Table~\ref{tab:gal_PA}). We find good agreement between the two independent analyse, with a median $\Delta$PA$_{gas}$\,$=$\,PA$^{CAMEL}_{gas}$\,--\,PA$^{kin}_{gas}$~of $\approx$\,7$^\circ$\,$\pm$\,31$^\circ$, and consistent measurements at less than 20$^\circ$ for 8/10 galaxies (see histogram on Fig.~\ref{fig:PA_gas_ppxf_camel}). Two galaxies, \id5 and
\id12, have a $\Delta$PA$_{gas}$ greater than 20$^\circ$ mainly because of the low spatial resolution of our maps (see Fig.~\ref{fig:kin_maps12}, panel {\it f}). The derived value of PA$^{kin}_{gas}$ is thus poorly constrained, which is captured by the large uncertainties on $\Delta$PA$_{gas}$ for these latter two galaxies. We observe a significant difference in the kinematics between the [OIII] doublet and the Balmer series for \id5 (both in PA and velocity amplitude) that could also partially explain the misalignment observed with~\citetalias{Contini2016}, who fitted the group (\protect{[OIII]}, H$\beta$) as one kinematic component. Despite the good agreement on, for example\, \id11, the large uncertainties visible on Fig.~\ref{fig:PA_gas_ppxf_camel} translate again the low spatial resolution of the kinematic maps.

The gas of the \udft~galaxies shows a velocity gradient of about 30\,--\,100\,\kms and a velocity dispersion of about 30\,--\,70\,\kms, which is typical of these intermediate redshift galaxies and similar to the HDFS galaxies (see \citetalias{Contini2016}). A dedicated analysis of the gas kinematics of the MUSE HUDF galaxies will be presented in \cprepp.

\begin{table*}
\caption{Kinematics results}
\label{tab:gal_PA}
\centering
\begin{tabular}{lrcccccc} 
\hline\hline\\
Field & ID & $\epsilon$ & PA$_{phot}$ & PA$^{kin}_{stellar}$ & PA$^{kin}_{gas}$ & S/N & Ref. line\\
- & - & - & ($^\circ$) & ($^\circ$) & ($^\circ$) & - & -\\ 
(1) & (2) & (3) & (4) & (5) & (6) & (7) & (8)\\
\hline\\
HDFS & 1 & 0.41 & 31.4$\pm$0.1 & 23.5$\pm$18.0 & 30.5$\pm$15.0 & 15 & [OIII] doublet\\
HDFS & 3 & 0.06 & -18.8$\pm$3.0 & 50.5$\pm$11.5 & 59.5$\pm$21.8 & 15 & Balmer series\\
HDFS & 4 & 0.44 & 35.8$\pm$0.1 & 48.0$\pm$7.5 & 48.0$\pm$14.5 & 15 & [OIII] doublet\\
HDFS & 5 & 0.17 & 6.9$\pm$0.2 & 14.0$\pm$16.5 & 32.0$\pm$35.3 & 15 & Balmer series\\
HDFS & 6 & 0.16 & 45.0$\pm$2.0 & 102.0$\pm$89.8 & 2.5$\pm$29.5 & 15 & [OIII] doublet\\
HDFS & 7 & 0.20 & 43.0$\pm$0.7 & 75.5$\pm$24.2 & 49.0$\pm$12.8 & 11 & [OIII] doublet\\
HDFS & 8 & 0.15 & -29.5$\pm$0.4 & -26.5$\pm$3.2 & -22.5$\pm$5.2 & 12 & Balmer series\\
HDFS & 9 & 0.19 & -34.8$\pm$0.4 & -36.0$\pm$7.5 & -36.0$\pm$9.8 & 15 & Balmer series\\
HDFS & 11 & 0.08 & 27.3$\pm$0.5 & -47.0$\pm$75.5 & 21.0$\pm$89.5 & 12 & [OIII] doublet\\
\medskip
HDFS & 12 & 0.42 & -144.0$\pm$4.9 & -125.5$\pm$41.8 & -103.5$\pm$89.8 & 12 & [OIII] doublet\\
UDF-10 & 1 & 0.09 & -35.7$\pm$0.1 & -30.0$\pm$6.5 & -28.0$\pm$4.8 & 15 & Balmer series\\
UDF-10 & 2 & 0.05 & 9.1$\pm$0.1 & -9.0$\pm$11.7 & -13.0$\pm$11.8 & 15 & Balmer series\\
UDF-10 & 3 & 0.20 & 52.4$\pm$0.1 & 51.0$\pm$8.2 & 56.5$\pm$6.8 & 12 & Balmer series\\
UDF-10 & 4 & 0.18 & 64.2$\pm$0.1 & 80.0$\pm$12.2 & 81.5$\pm$11.5 & 12 & Balmer series\\
UDF-10 & 5 & 0.07 & 31.4$\pm$0.1 & 17.0$\pm$89.8 & - & 12 & -\\
UDF-10 & 7 & 0.16 & 43.2$\pm$0.1 & -68.0$\pm$89.8 & -14.0$\pm$89.8 & 12 & [OIII] doublet\\
UDF-10 & 10 & 0.19 & -5.8$\pm$0.1 & 63.0$\pm$10.2 & -8.0$\pm$24.3 & 10 & Balmer series\\
\hline
\end{tabular}
\tablefoot{Kinematics parameters of the galaxy sample. (1) Field of observations (2) galaxy's ID; (3)  ellipticity of galaxy measured on the MUSE white light image, at 2\,R$_{e}$, and derived as the first moment of the surface brightness; (4) photometric major-axis position angle (PA); (5) stellar kinematic major-axis PA; (6) gas kinematics major-axis PA; (7) targeted spectral S/N of the stellar continuum of the binned sub-cubes; and (8) brightest emission line detected in the MUSE sub-cubes, used to constrain gas kinematics.}
\end{table*}

\subsection{Resolved stellar kinematics}
\label{subsec:kin_results}

The resolved stellar kinematic maps of our galaxy sample are presented in Fig.~\ref{fig:Main_kin_fig} for the \idu1 galaxy and in Figs.~\ref{fig:kin_maps1} to \ref{fig:kin_maps10_udf} for the other galaxies. The values quoted here refer to what we observed within the spatial coverage of the MUSE data, given the magnitude limit of the cube and our S/N rejection criterion of the spectra (see \S\ref{subsec:binning}).

Ten galaxies show clear stellar rotation (\id1, \id3, \id4, \id5, \id8, \id9, \idu1, \idu2, \idu3, and \idu4) with a maximum amplitude ranging from $\pm$40\,--\,130\,\kms. The maximum stellar rotation amplitude is observed for the two almost edge-on disk-like galaxies (\id4 at 130\,\kms and \id9\,at\,100\,\kms), whereas the rather face-on spiral galaxies (\id3, \idu1, \idu2, and \idu4) show lower stellar rotation ($\pm$40\,--\,80\,\kms). Nevertheless, it is worth noticing that all galaxies showing large amplitude stellar rotation ($\geq$\,40\,\kms) are either edge-on disk-like galaxies or exhibit a spiral morphology. Four galaxies show low stellar rotation given our kinematics uncertainties (\id7, \id12, \idu5, and \idu7) with a maximum radial velocity of $\approx$ 20\,--\,30\,\kms. Their rotational pattern also appears less regular, which is probably due to pair interactions (\id7 and \id12; see \citetalias{Contini2016}), clumpy morphology (\idu7), or simply the combination of low spatial resolution and relatively low S/N. Finally, three galaxies show very little to no stellar rotation pattern (\id6, \id11, and \idu10) with radial velocity amplitudes lower than $\approx$\,15\,\kms. These three galaxies are either compact (\id11, and therefore the spatial resolution is very poor), or exhibits strongly perturbed morphology including streams, satellites (\idu10), or are seen at low inclinations (\id6; e.g\, $i \approx 29^\circ$).

Regarding the stellar velocity dispersion, the majority of the galaxy sample show flat resolved maps (given the large kinematics uncertainties) except for two galaxies (\idu2 and \idu5) that show centrally peaked stellar velocity dispersion maps. We measured for these two galaxies, $\sigma_{stellar}$\,$\approx$\,40\,--\,65\,\kms~and $\sigma_{stellar}$\,$\approx$\,110\,--\,180\,\kms, respectively, from their outskirts to their centres. Five galaxies have low stellar velocity dispersion (\id1, \id3, \id6, \id7, and \idu7) of $\approx$30\,--\,40\,\kms~and nine galaxies (\id4, \id5, \id8, \id9, \id11, \id12, \idu1, \idu3, and \idu4) show higher values between $\approx$50\,--\,90\,\kms. Finally, \idu10 shows very large spatial variations in its stellar velocity dispersion map, most probably due to the low S/N (i.e.\, 10\,\ppix) and low spatial resolution of the galaxy binned data cube.

Overall, the observed stellar kinematics of the intermediate redshift galaxies seem to have patterns that are consistent with their respective gas kinematics, both in terms of the mean rotation and velocity dispersion. This nicely supports the underlying morphology revealed by their high-resolution HST images. Indeed, the clear spirals and disk-like galaxies (\id1, \id3, \id4, \id5, \id6, \id9, \idu1, \idu2, \idu3, \idu4, and \idu5) mostly show regular rotation with significant amplitude whereas the more compact objects (\id11, and \id12) and morphologically disturbed (\id7, \id8, \idu7, and \idu10) show lower rotation, with higher velocity dispersion sometimes, both for gas and stellar components.

More interestingly, these observations of the resolved stellar kinematics of intermediate redshift galaxies ($0.2 \leq z \leq 0.8$) show similar patterns as observed in disk-like galaxies of the local Universe~\citep[i.e.\,][]{Falcon-Barroso2016}, which would suggest that the \textit{regular} stellar kinematics of galaxies that we see in the local Universe was already in place 4\,--\,7\,Gyr ago.\\

For illustration purpose, we also present in appendix~\ref{sec:appendixB} the one-dimensional velocity and velocity dispersion fields of the individual sample galaxy; see Fig.~\ref{fig:rot_curve_hdfs} and Fig.~\ref{fig:rot_curve_udf}.

\section{Discussion}
\label{sec:dicussion}

\subsection{Kinematics misalignment angles}
\label{subsec:kin_PA}

We measured the stellar kinematics major-axis PA (PA$^{kin}_{stellar}$), and the gas kinematics major-axis PA (PA$^{kin}_{gas}$) from the two-dimensional maps presented in sections \S\ref{subsec:gas_kin} and \S\ref{subsec:kin_results}, using the method presented in \cite{Krajnovic2006} again. Both obtained quantities are listed in Table~\ref{tab:gal_PA} for the whole galaxy sample. The morphological, stellar, and gas kinematics major-axis PAs of the galaxy sample are on average aligned and the observed differences can be mostly explained by a combination of the galaxy inclinations (ellipticities), morphological features (such as bars and clumps), and the low spatial resolution of the kinematic maps. This suggests that the observed galaxies are disk dominated, oblate, and axisymmetric systems, allowing for perturbations such as bars and spiral arms, as observed in the deep HST images.

\begin{figure*}
        \includegraphics[width=\textwidth]{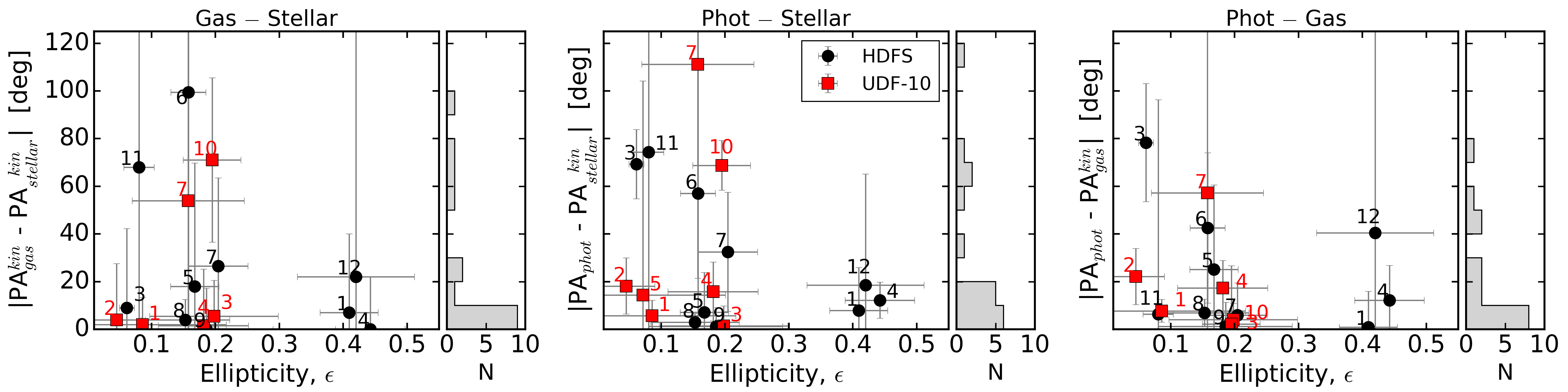}
        \caption{Comparison of the gas and stellar kinematics major-axis PA (left panel), the photometric major-axis PA (PA$_{phot}$) and the stellar kinematics PA$^{kin}_{stellar}$ (middle panel), and the PA$_{phot}$ with the gas kinematics PA$^{kin}_{gas}$ (right panel). The distributions of the kinematics misalignment angles (KMA) are shown on the right side of each plot. Despite the large uncertainties, we find a global alignment of the morphological, stellar, and gas kinematics major-axis PAs. The galaxy ID numbers are indicated above each point.}
    \label{fig:PA_all}
\end{figure*}

In support of these claims, we first compare, in Fig.\,\ref{fig:PA_all} (left panel), the stellar kinematics PA$^{kin}_{stellar}$ to the gas kinematics PA$^{kin}_{gas}$, as a function of the ellipticity of the galaxy. We find that for the majority of the sample (11/17), the stellar and gas kinematics major axes are aligned within 20$^\circ$. The median stellar\,--\,gas misalignment is 9$^\circ$ with a standard deviation of 29$^\circ$. Two galaxies have larger values of $\approx$\,20 to 30\,$^{\circ}$ and four have values between 40 and 100\,$^{\circ}$ (see left histogram on Fig.\,\ref{fig:PA_all}), although are all consistent with zero. All these seven galaxies have low spatial resolution kinematic maps (i.e.\, a low number of spatial bins) and noisy kinematic maps (see respective figures in appendix~\ref{sec:appendixA}) that most likely explain the observed departures. These effects are captured by the large uncertainties on the PA$^{kin}_{stellar}$ and PA$^{kin}_{gas}$ measurements and the largest stellar\,--\,gas misalignment found mostly at low ellipticity (i.e.\, face-on galaxies and poor spatial resolution). For a few galaxies, for example \id6, \idu7, and \idu10, the stellar\,--\,gas misalignment could come from their asymmetric/perturbed morphology,  possibly triggered by galaxy\,--\,galaxy interactions (see also \citealt{Davis2011a, Barrera2015}), but the large uncertainties on the measurements do not allow us to confirm such a statement. Overall, accounting for the large uncertainties on the PA measurements of the stellar and gas kinematics major axes, the results point towards a global alignment of the two components, as expected for disk-like galaxies.\\

We also compare in Fig.\,\ref{fig:PA_all} (middle panel) the stellar kinematics PA$^{kin}_{stellar}$ to the photometric major-axis PA (PA$_{phot}$). The latter has been derived in~\citetalias{Contini2016} for the HDFS galaxies and in \Inami~for the \udft~galaxies. The difference between the two quantities, $\Delta$PA = $|$PA$_{phot}$\,-\,PA$^{kin}_{stellar}|$ is a positive value between 0$^\circ$ and 90$^\circ$, called the stellar {\it kinematics misalignment angle} (KMA), as defined in~\citet{Franx1991}. We find that the stellar and photometric major axes are well aligned within 20$^\circ$ for 11/17 galaxies and that the whole galaxy sample has a median stellar KMA of $\approx$\,16$^\circ$ with a standard deviation of 32$^\circ$. Four galaxies show a significantly larger stellar KMA of $\approx$\,60\,--70\,$^\circ$: \id3, \id6, \id11, and \idu10, all of these being either face-on or showing morphological features (such as bar, clumps) that strongly affect the accuracy of the measured photometric PA$_{phot}$. \idu7 has a large stellar KMA (kept larger than 90$^\circ$ in order to minimise the gas KMA; see next paragraph) that results from the combination of the low spatial resolution of its kinematic maps and its strong bar visible in the HST image. Once again, this is captured by the large uncertainties on the measured PA$^{kin}_{stellar}$. 

Finally, we compare the gas kinematics PA$^{kin}_{gas}$ to the photometric major-axis PA, PA$_{phot}$ (see Fig.~\ref{fig:PA_all}, right panel), and find that the gas kinematic major axis is aligned within 20$^\circ$ for 10 of 17 galaxies with the photometric major axis. One galaxy (UDF10-ID5) has no ionised gas and is excluded from the plot. The median gas KMA is 10$^\circ$ with a standard deviation of 22$^\circ$ that is consistent with results of \citetalias[][see their Fig.~9]{Contini2016}, and as expected from the near kinematic alignment between stars and gas and the low KMA.

\subsection{Stellar and gas kinematics}
\label{subsec:stellar_gas_kin}

We have seen  the close similarities of the stellar and gas kinematics of the intermediate-redshift galaxy sample in \S\,\ref{subsec:kin_results}. As a first step towards a better quantification of these similarities, we now present a direct comparison of the second order velocity moment between the stellar and gas components that we approximate as $V_{\rm rms}~=~\sqrt{V^2 + \sigma^2}$, where $V$ and $\sigma$ are the mean velocity and velocity dispersion of the respective component measured along the line of sight for each spatial bin.

\begin{figure}
        \centering
        \includegraphics[width=\columnwidth]{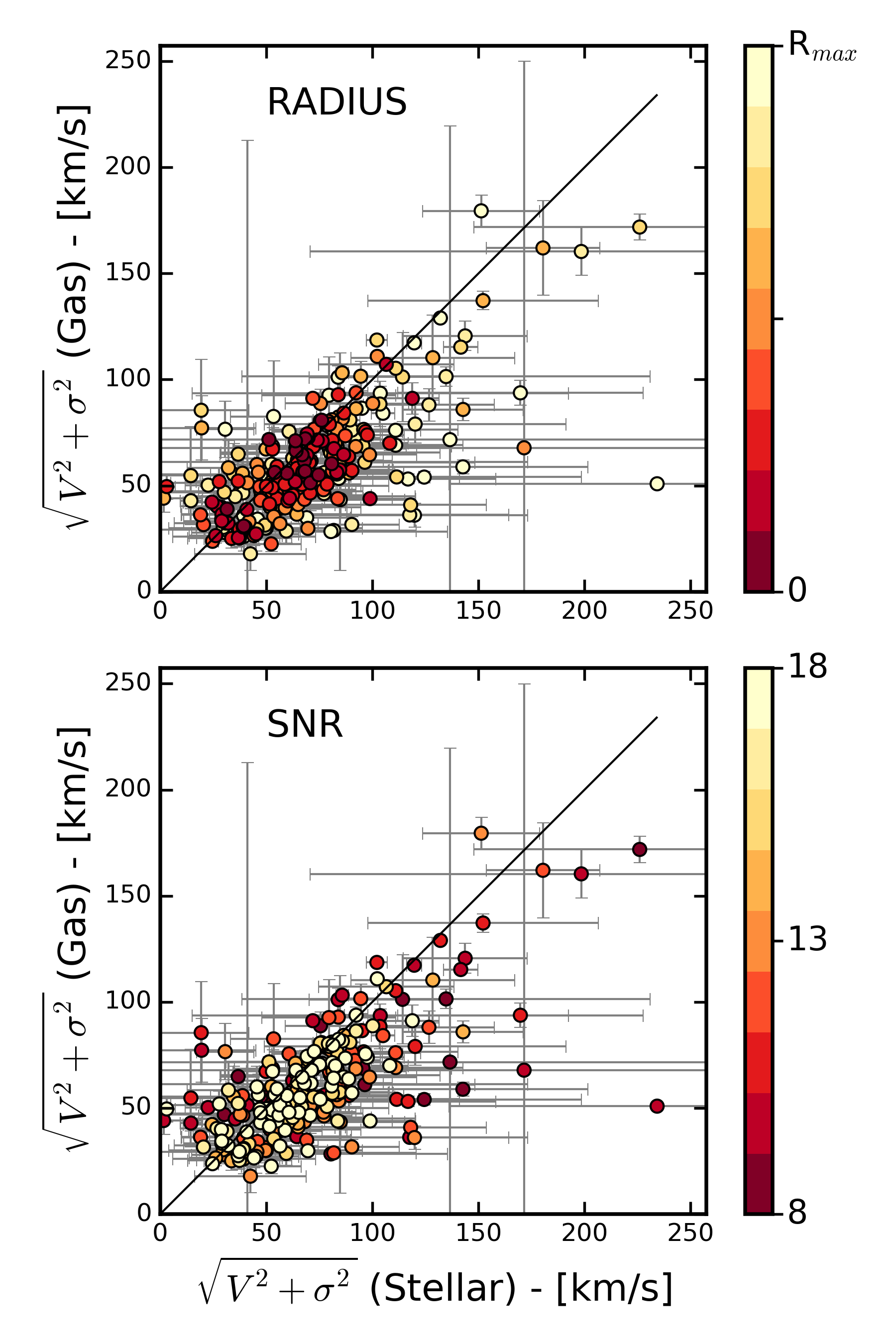}
        \caption{Second order velocity moment of the stellar and gas components, $V_{\rm rms}=\sqrt{V^2 + \sigma^2}$, for all the galaxies of the sample. Each point corresponds to a spatial bin of MUSE sub-cubes and is colour coded by its respective radial distance (top panel) and its respective S/N bin (bottom panel). The 1:1 relation is indicated by the black lines.} 
        \label{fig:v2s2}
\end{figure}

Indeed, this quantity scales, in principle, with the underlying gravitational potential of the galaxy in the situation where the kinematics is driven mostly by gravity (and not by, for example turbulent motions, heating, cooling, and shocks). On the contrary, if non-gravitational forces dominate, the gas velocity dispersion increases significantly because of the dissipational character of gas interactions, while the stellar component is only mildly affected. As a consequence, under the assumption that the gas and stellar spatial distributions are similar, we expect the $V_{\rm rms}$ of the gas to be systematically higher than that of the stars when strong non-gravitational forces act on the gas. A direct comparison of the stellar and gas $V_{\rm rms}$ can thus partially address the origin of the gas velocity dispersion, which is a debated topic for intermediate- to high-redshift galaxies~\citep{Epinat2010, Epinat2012camel, Green2010, Lehnert2013, Newman2013, Buitrago2014,Ubler2017}. This direct comparison is shown in Fig.\,\ref{fig:v2s2}, where we plot $V_{\rm rms}^{gas}$ as a function of $V_{\rm rms}^{stellar}$ for each spatial bins of the kinematic maps of all the galaxies in the sample. The top panel shows each bin colour coded as a function of its radial position and the bottom panel as a function of the respective bin S/N.

Fig.\,\ref{fig:v2s2} shows that the $V_{\rm rms}$ of the gas and stellar components seem to follow a 1:1 relation with a 1$\sigma$ standard deviation of $\approx$\,25\,\kms. Clear trends with the radial distance and the S/N of the spatial bins are visible: bins at larger radial distances (see top panel) and with a relatively low S/N (see bottom panel) are on average further away from the 1:1 relation. On the contrary, the bins with relative high S/N and located at the centre of the galaxies on average follow the 1:1 relation well. This would thus suggest that the deviation from the 1:1 relation is mostly due to the measurements uncertainties. Furthermore, the scatter of the distribution increases for large values of $V_{\rm rms}$ ($\geq$\,100\,\kms), where $V_{\rm rms}^{stellar}$ is systematically higher than $V_{\rm rms}^{gas}$. We interpret this as the consequence of some ``unreliable'' measurements of $\sigma_{stellar}$ in the outskirts of the MUSE sub-cubes when the S/N is low (i.e.\, outliers bins with high measured values of $\sigma$; see e.g.\, \id8, \id9, \idu4, and \idu7). Finally, the low spatial resolution of the kinematic maps certainly impacts the observed trends, i.e.\, the beam smearing is not taken into account.

Despite the large uncertainties of some of these measurements, Fig.\,\ref{fig:v2s2} suggests that the $V_{\rm rms}$ of the stellar and gas components are roughly comparable and thus that the observed gas kinematics traces the gravitational potential of the intermediate-redshift sample galaxies (as more robustly traced by the stellar kinematics). The gas kinematics of the galaxy sample is thus likely not dominated by shocks and turbulent motions. This result should be confirmed with a set of full-fledged dynamical models, which we briefly touch upon in the following section.

\subsection{Examples of dynamical mass estimates}
\label{subsec:dyn}

Spatially resolved stellar kinematics of galaxies can be used to constrain dynamical models, including the dark matter (DM) content of galaxies, and to determine their dynamical masses. The MUSE data of intermediate redshift galaxies are of sufficient quality to constrain more sophisticated dynamical models than previously done based on their stellar components. In this section, we present dynamical models based on the stellar kinematics for two galaxies, \id4 and \idu1, which we selected as they have among the best spatial resolutions and best data quality of the galaxy sample, which makes them the best-suited objects for a case by case study. A more detailed investigation of the dynamical properties of the full sample will follow in a separate paper. 

\subsubsection{Building of the dynamical models}
\label{subsubsec:dyn_model}

To build the dynamical models, we first parametrise the HST images in the F814W filter using the Multi-Gaussian Expansion (MGE) method \citep{Emsellem1994}, in the implementation of \citet{Cappellari2002}\,\footnote{\url{www-astro.physics.ox.ac.uk/~mxc/software/\#mge}}. Assuming a mass-to-light ratio (M/L), an inclination, and an axisymmetric shape for the galaxy, one can deproject its MGE model into a three-dimensional mass density \citep{Monnet1992}, which defines the potential and distribution of the tracers (stars or gas) within the Jeans Anisotropic Modelling (JAM; \citealt{Cappellari2008Jam})\,\footnote{\url{www-astro.physics.ox.ac.uk/~mxc/software/\#jam}} scheme. Here we consider self-consistent models, without dark matter, and thus explicitly assume that mass follows light; the contribution of dark matter is included in the final M/L ratio.

The best-fit models are constrained by the observed second order velocity moment of the stars, $V_{\rm rms}~=~\sqrt{V^2 + \sigma^2}$, where again, $V$ and $\sigma$ are the mean velocity and velocity dispersion of the stellar component measured along the line of sight for each spatial bin. The shape and scale of this quantity are specified in the models by three parameters that are varied to find the best fitting model: the inclination, M/L, and the velocity anisotropy $\beta_z = 1- \sigma_z^2/\sigma_R^2$. We estimated the errors on $V$ and $\sigma$ through Monte Carlo simulations (see section \S\,\ref{subsec:binning}) and used the standard propagation of errors formula to estimate the errors on $V_{\rm rms}$. However, because errors increase by a factor of~3\,--\,5~from the centre to the outskirts of the galaxies, to avoid biasing the solutions by a few central spatial bins, we 
assumed an average error for all spatial bins for $V$ and for $\sigma$ and used these values in the formula for the propagation of the errors. This means that we obtain nearly constant errors on $V_{\rm rms}$ across the full map.

As shown in Table~\ref{tab:gal_prop}, the inclinations of \id4 and \idu1 determined from their HST images are 75\degree and 27\degree, respectively. We ensure that the MGE models allow for such inclinations and then explore the full parameter space (M/L, i, $\beta_z$) using the {\tt emcee} Markov Chain Monte Carlo sampling code \citep{Foreman-Mackey2013}, an implementation of the \citet{Goodman2010} sampler. In our runs, we use an uninformative prior function and a likelihood function based on a Gaussian distribution of errors. This implementation is similar to that of \citet{Mitzkus2016}, specifically their model {\it a}.\\

The obtained best-fit parameters (M/L, i, $\beta_z$) of the respective dynamical models are given in Table~\ref{tab:models}.\\

\begin{table}
\caption{JAM models}
\label{tab:models}
\centering
\begin{tabular}{cccc} 
\hline\hline\\
Galaxy& M/L & {\rm \it{i}} & $\beta_z$ \\
-     &  -  & ($^\circ$) &     -     \\
(1) & (2) & (3) & (4)\\
\hline\\
\id4 & $6.09^{+0.3}_{-0.4}$ & $85^{+4}_{-4}$ & $0.05^{+0.14}_{-0.19}$\\
\\
\idu1 & $1.58^{+0.12}_{-0.06}$ & $26^{+16}_{-0.05}$ & $0.1^{+0.16}_{-0.09}$ \\
\hline
\end{tabular}
\tablefoot{Parameters of the best-fit JAM models for \id4 and \idu1. (1)~Galaxy's ID; (2)-(3)-(4)~mass-to-light ratio (in the V band), inclination, and velocity anisotropy derived from the JAM models}
\end{table}

\subsubsection{M/L and dynamical masses}
\label{subsubsec:dyn_model_results}

From the above described dynamical models, we obtain M/L ratios of 6.09 for \id4, and 1.58 for \idu1, which correspond to typical M/L ratios in the V band\,\footnote{The F814W HST filter roughly corresponds to the V band at these redshifts of $z\approx$\,0.6}, both in the local Universe~\citep{Bell2001} and for intermediate-redshift galaxies~\citep[][based on long-slit data]{Shetty2015a}. The implied total dynamical masses using the luminosity derived from the HST photometry and the assumption that mass follows light would represent massive spirals (i.e.\, $\sim$\,10\,$^{11}$\,\Msun\,; see Table~\ref{tab:mass_results}). In comparison, the stellar masses obtained from SED fitting are significantly lower  (i.e.\, $\sim$\,10\,$^{9\,-\,10}$\,\Msun\,, see Table~\ref{tab:mass_results}). If one assumes an average gas mass of $\sim$\,35\% of the baryonic mass for these intermediate redshift galaxies~\citep{Stott2016}, one obtains a fraction of DM $f_{DM}$~$\sim$\,80\% for \id4, and $f_{DM}$~$\sim$\,50\% for \idu1, hence hinting at DM dominated galaxies within a few effective radii. These results are consistent with the findings of~\citet[][based on the gas kinematics]{Stott2016}, but in disagreement with the findings of~\citet[][based on the stellar kinematics from long-slit data and stellar-to-halos mass abundance matching technique]{Shetty2015a} who found that their sample of 154 galaxies at redshift $z\approx0.8$ are not DM dominated within one effective radius ($f_{DM}$ $\approx$\,10\%). Our result is however to be taken with a lot of care. Indeed, we show here only two galaxies, we do not include DM halos in the mass models, we do not probe the same scale (a few R$_e$ versus one), and we would need to more accurately estimate the stellar and gas masses from the MUSE spectra directly.\\ 

Up to now, dynamical masses of intermediate-redshift galaxies were mostly derived through the use of the observed gas kinematics, based on either its rotation field only {\bf and} taking into account the asymmetric drift~\citep{ForsterSchreiber2009, Price2016, Stott2016, Pelliccia2017}. The gas component of intermediate-redshift galaxies could however be subject to non-gravitational forces (i.e.\, heating, cooling, and shocks) that weaken the different conclusions of these latter results. The stellar component of galaxies are less affected by such forces and it thus seems natural to directly compare the dynamical masses obtained from models based on the gas kinematics to the now available two-dimensional stellar kinematics data.   

Both of the stellar and gas components of \id4 and \idu1 rotate regularly. Their gas velocity dispersion maps are overall low ($\leq$\,60\,\kms) with a central peak and decrease towards lower values in the outskirts ($\approx$\,30-40\,\kms). As a consequence, and as a first approximation, we build simple dynamical models based on the gas rotation only (i.e.\, not accounting for the gas velocity dispersion). A simple order of magnitude calculation of the dynamical mass is then given by\begin{equation}
M_{\rm dyn}^{G}(R)\,=\,V_G^2\,R\,/\,G
\label{eq:mdyn}
,\end{equation}

where $V_G$ is the rotational velocity of the gas in the plane of the galaxy at the radius $R$, and $G$ is the gravitational constant. To estimate $V_G$, we used the parametric model of the gas rotational field of \id4 described in~\citetalias{Contini2016}, and in \cprepp~for \id1. Both models use a parametrisation of the rotation curve corresponding to an exponential disk and are based on the ``high'' resolution kinematic maps of the gas component (see \citetalias{Contini2016}).  We used the inclinations obtained from the HST images (see Table~\ref{tab:gal_prop}) to deproject the observed gas rotation velocity fields on the respective galaxy's plane. 

Using eq.\,(\ref{eq:mdyn}), we estimated the dynamical mass of both galaxies at one effective radius, R$_e$, such that it is well covered by the MUSE data, hence the kinematics is well constrained by the observations. At the respective redshifts, and with our choice of cosmology, the effective radius of \id4 is $R_e=8.9$ kpc, while \idu1 has $R_e = 8.5$ kpc. We obtain $V_G$\,(R$_e$)\,=\,175\,\kms~for \id4, and $V_G$\,(R$_e$)\,=\,168\,\kms~for \idu1, and thus, a dynamical mass within one effective radius, M$_{\rm dyn}^G$(R$_e$), of $6.3^{+0.3}_{-0.2}\times10^{\,10}$\,M$_\odot$ for \id4, and $5.6^{+2.6}_{-1.5}\times10^{\,10}$\,M$_\odot$ for \idu1 (see Table~\ref{tab:mass_results}). We then derived the dynamical mass within one R$_e$ from the JAM models (see \S\,\ref{subsubsec:dyn_model}), which are based on the stellar kinematics and obtain M$_{\rm dyn}^S$(R$_e$)\,=\,$5.7^{+0.2}_{-0.4}\times10^{\,10}$\,M$_\odot$ for \id4, and $4.5^{+0.3}_{-0.2}\times10^{\,10}$\,M$_\odot$ for \idu1 (see Table~\ref{tab:mass_results}).\\

\begin{table}
\caption{Dynamical mass estimates}
\label{tab:mass_results}
\centering
\begin{tabular}{ccccc} 
\hline\hline\\
Galaxy & \Mstar & M$_{\rm dyn}^{S}$ & M$_{\rm dyn}^{S}$(R$_e$) & M$_{\rm dyn}^{G}$(R$_e$)\\
\medskip
- & (1) & (2) & (3) & (4) \\
\hline\\
\id4 & $0.72^{+0.35}_{-0.23}$ & $10.2^{+0.05}_{-0.06}$ & $5.7^{+0.2}_{-0.4}$ & $6.3^{+0.3}_{-0.2}$\\
\\
\idu1 & $2.75^{+0.48}_{-0.56}$ & $14.9^{+0.1}_{-0.07}$ & $4.5^{+0.3}_{-0.2}$ & $5.6^{+2.6}_{-1.5}$ \\
\\
\hline
\end{tabular}
\tablefoot{Mass estimates (in units of [$10^{10}$\,\Msun]) for \id4 and \idu1. (1) Total stellar mass measured in~\citetalias{Contini2016}, and \cprepp using SED fitting; (2) total dynamical mass derived from the stellar kinematics using JAM models; (3) dynamical mass within one effective radius derived from the stellar kinematics using JAM models, M$_{\rm dyn}^{S}$; and (4) dynamical mass within one effective radius derived from the gas kinematics modelling, M$_{\rm dyn}^{G}$.}
\end{table}

Despite the higher values for M$_{\rm dyn}^G$(R$_e$) than M$_{\rm dyn}^S$(R$_e$), these values compare well within 10\,--\,25\%. This follows the naive expectation provided by the fact that the second order velocity moments estimated from the gaseous and stellar components compare well to each other (see\, Fig.~\ref{fig:v2s2}). Obviously, these estimates should be refined, considering the actual spatial extent and detailed distribution of the tracers, the respective assumptions behind such derivations, and due to limitations such as the existing degeneracies at high inclinations of the JAM models, the non-inclusion of DM haloes, and the gas velocity dispersion that is not accounted for.

\section{Summary}
\label{sec:summary}

We have presented two-dimensional, spatially resolved, stellar kinematics of galaxies at intermediate redshift ($0.2 \lesssim z \lesssim 0.8$). We used the two deepest MUSE observations to date ($\approx$\,30\,h exposure per field) in the Hubble Deep Field South (HDFS) and Hubble Ultra Deep Field-10 (\udft) to select a sample of 17 galaxies with spatially resolved stellar continuum. Combined with high-resolution deep HST images, we show that this sample is mainly composed of late-type disk, main-sequence star-forming galaxies with \Mstar\,$\sim$\,1$0^{\,8.5\,-\,10.5}$\,\Msun. One galaxy deviates from the normal star-forming sequence and this galaxy is suggested to be an early-type, disk-like, quiescent field galaxy. Thanks to a careful determination of the MUSE spectral resolution, we derived maps of the rotational velocity, $V$, and the velocity dispersion, $\sigma$, for both the stellar and gas components. These unique two-dimensional maps allowed us to unveil new insights on the baryonic kinematic properties of intermediate redshift galaxies.

The observed stellar kinematics of most of the galaxy sample is consistent with a rotating stellar disk, as first expected from their already available photometry and gas kinematics. The amplitude of the stellar rotation ranges from almost null for face-on, perturbed galaxies, to $\approx$\,40\,--\,130\,\kms~for disk-like edge-on galaxies. The associated stellar velocity dispersions are rather constant over the spatial coverage for all galaxies in the sample and range from $\approx$\,30\,--\,90\,\kms. Despite the limited spatial resolution of the MUSE cubes and relatively low S/N of the stellar continuum, the derived stellar kinematic maps suggest that the {\it regular} stellar kinematics that is observed in the local Universe was already in place 4\,--\,7\,Gyr ago. 

We compared the position angle (PA) of the major axes of the derived stellar and gas kinematics. We found a median misalignment of 9$^\circ$\,$\pm$\,33$^\circ$ between the two components, suggesting a global alignment of the stellar and gas kinematics rotation axes, as expected for disk-like galaxies. We also compared the PA of the stellar kinematics and photometric major axes, so-called {\it kinematics misalignment angle}, and found a median misalignment of 16$^\circ$\,$\pm$\,14$^\circ$. This suggests that the galaxy sample is dominated by oblate, axisymmetric systems. The observed differences between the stellar kinematics major axis and both, the gas kinematics and photometric major axes, can mostly be explained by a combination of the inclination, perturbed morphology, and the limited spatial resolution of the kinematic maps of the galaxy.

Thanks to the spatially resolved kinematic maps, we studied the second velocity moment, $V_{\rm rms}=\sqrt{V^2 + \sigma^2}$, of both the stellar and gas components of the 17 intermediate redshift galaxies. We compared at faced values the derived $V_{\rm rms}$ in each spatial bins of the MUSE cubes between the two components and found that the galaxy sample approximately follows a 1:1 relation (Fig.\ref{fig:v2s2}). Because the second velocity moment of the stellar component is a robust tracer of the gravitational potential, the derived values suggest that the gas kinematics of the intermediate redshifts sample galaxies is indeed a good tracer of their gravitational potential and that their gas content is not dominated by turbulences.

Finally, we compared the dynamical masses of the two best-suited galaxies for a case by case study inferred from dynamical models. This study was based on the gas kinematics, on one hand, assuming simple exponential disks and the currently available two-dimensional stellar kinematics, on the other hand, based on axisymmetric dynamical models constrained by the observed second velocity moment. We found a reasonable agreement between the two methods, within 10\,--\,25\,\% in one effective radius, consistent with the above-mentioned comparison made on $V_{\rm rms}$. The dynamical modellings also provide us with total dynamical masses, and when compared to the stellar masses obtained through SED fitting, suggest that the two example galaxies are DM dominated within a few effective radii.

While these two cases already reveal the fantastic potential of the MUSE Ultra Deep Field data set, a more comprehensive study via refined dynamical models of the full sample (including a dark matter component) will be the topic of a forthcoming publication.

\FloatBarrier

\begin{acknowledgements}
This work has been carried out thanks to the support of the ANR FOGHAR (ANR-13-BS05-0010-02), the OCEVU Labex (ANR-11-LABX-0060), and the A*MIDEX project (ANR-11-IDEX-0001-02) funded by the ``Investissements d'avenir'' French government programme managed by the ANR. BE acknowledges financial support from the ``Programme National de Cosmologie et Galaxies'' (PNCG) of CNRS/INSU, France. RB acknowledges support from the ERC advanced grant MUSICOS. JR acknowledges support from the ERC starting grant CALENDS-336736. JS acknowledges support from the ERC grant 278594-GasAroundGalaxies. RAM acknowledges support by the Swiss National Science Foundation.

\end{acknowledgements}

\bibliographystyle{aa}

\begin{appendix}

\section{Stellar kinematic maps}
\label{sec:appendixA}

\begin{figure*}
        \includegraphics[width=\textwidth]{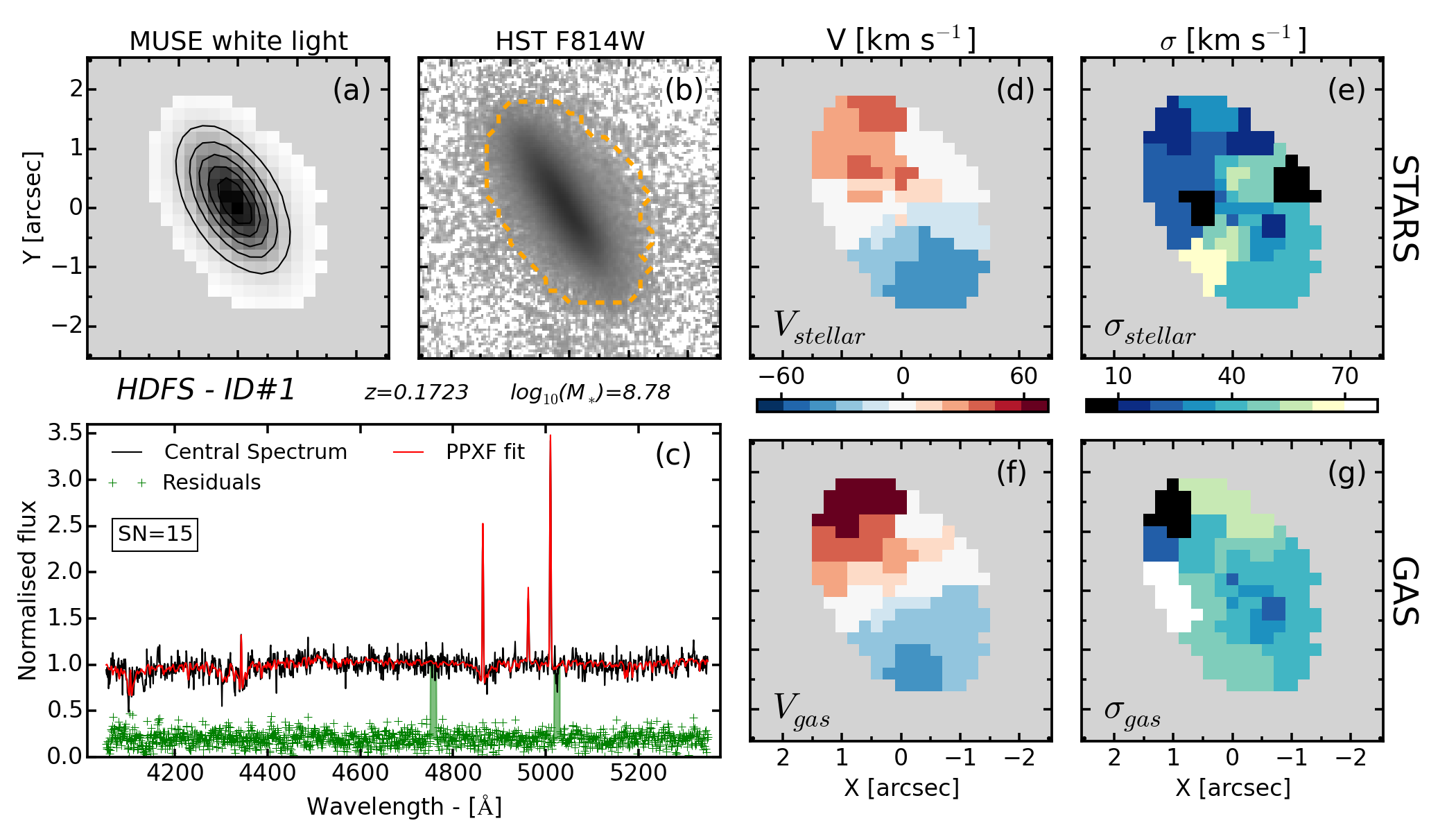}
        \caption{\id1 - caption as Fig.~\ref{fig:Main_kin_fig}}
        \label{fig:kin_maps1}
\end{figure*}

\begin{figure*}
        \includegraphics[width=\textwidth]{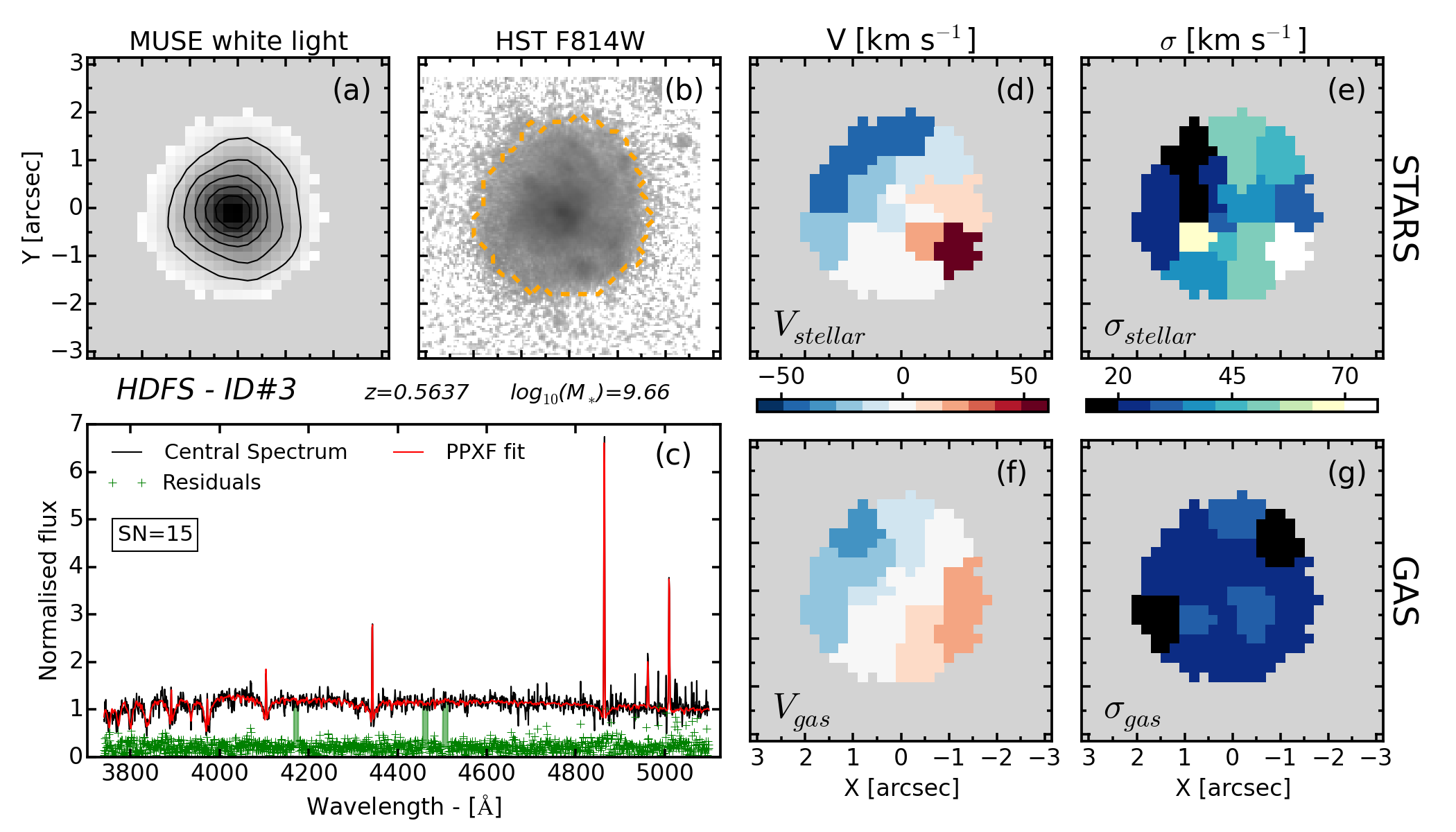}
        \caption{\id3 - caption as Fig.~\ref{fig:Main_kin_fig}}
        \label{fig:kin_maps3}
\end{figure*}

\begin{figure*}
        \includegraphics[width=\textwidth]{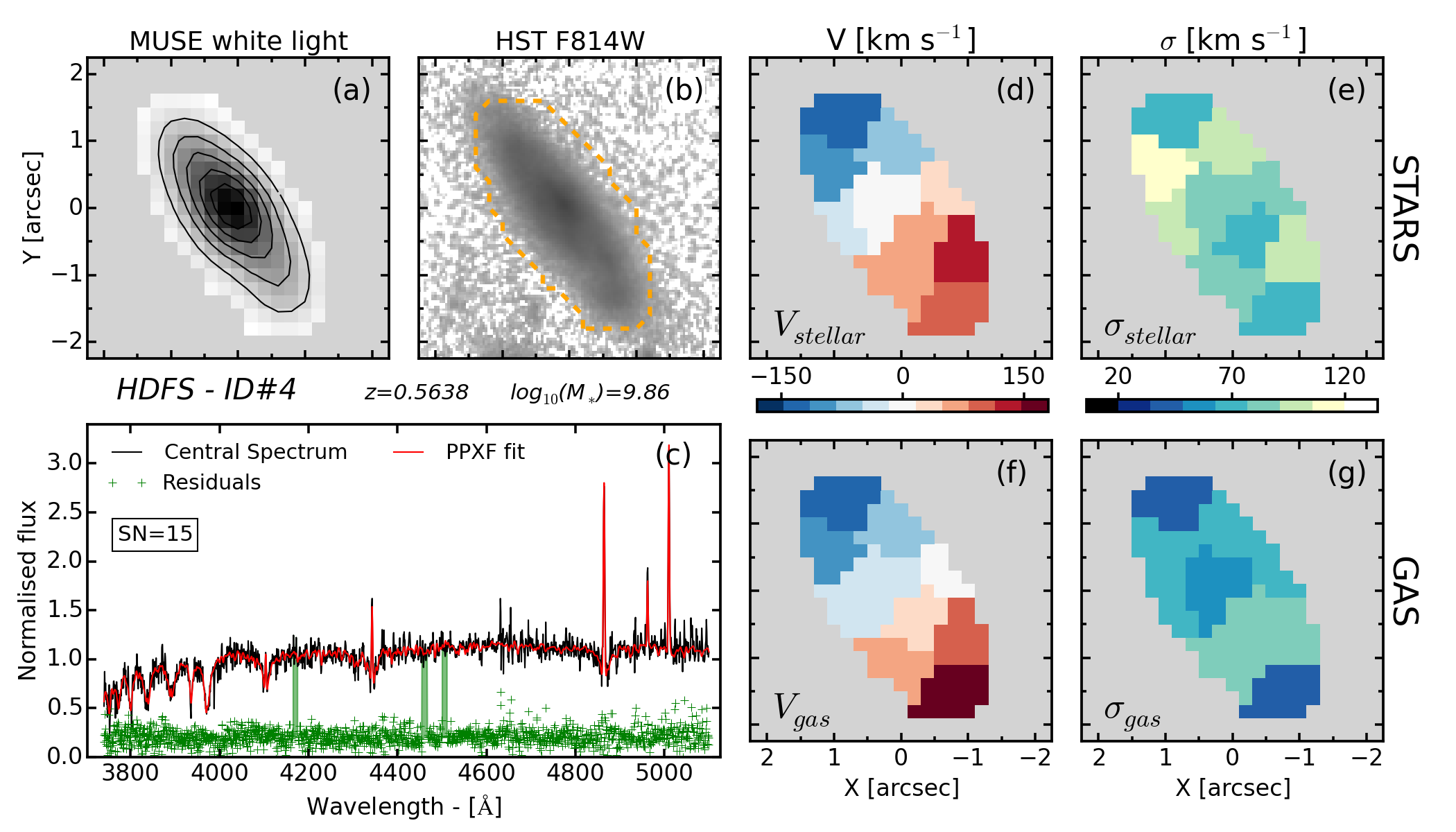}
        \caption{\id4 - caption as Fig.~\ref{fig:Main_kin_fig}}
        \label{fig:kin_maps4}
\end{figure*}

\begin{figure*}
        \includegraphics[width=\textwidth]{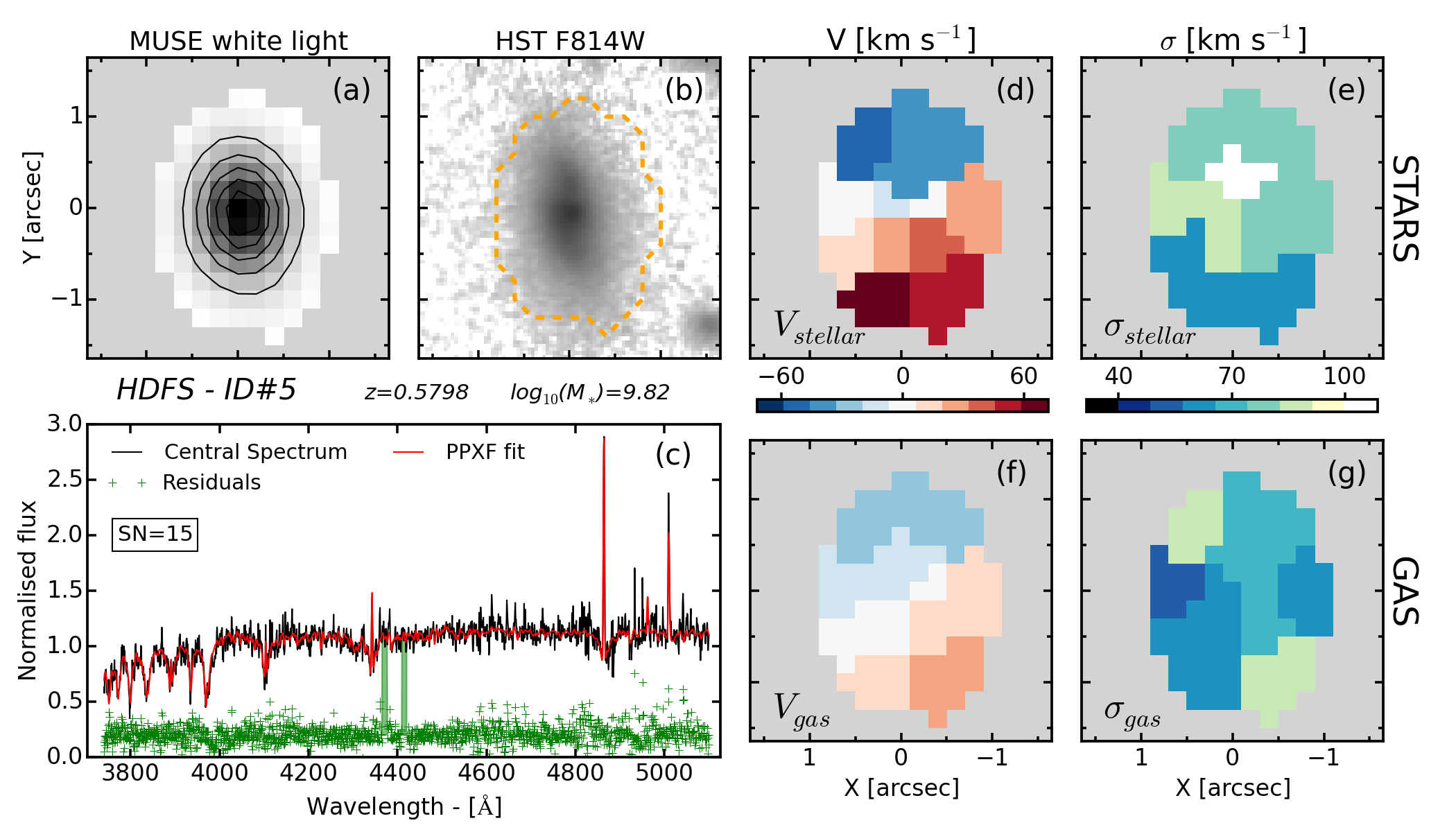}
        \caption{\id5 - caption as Fig.~\ref{fig:Main_kin_fig}}
        \label{fig:kin_maps5}
\end{figure*}

\begin{figure*}
        \includegraphics[width=\textwidth]{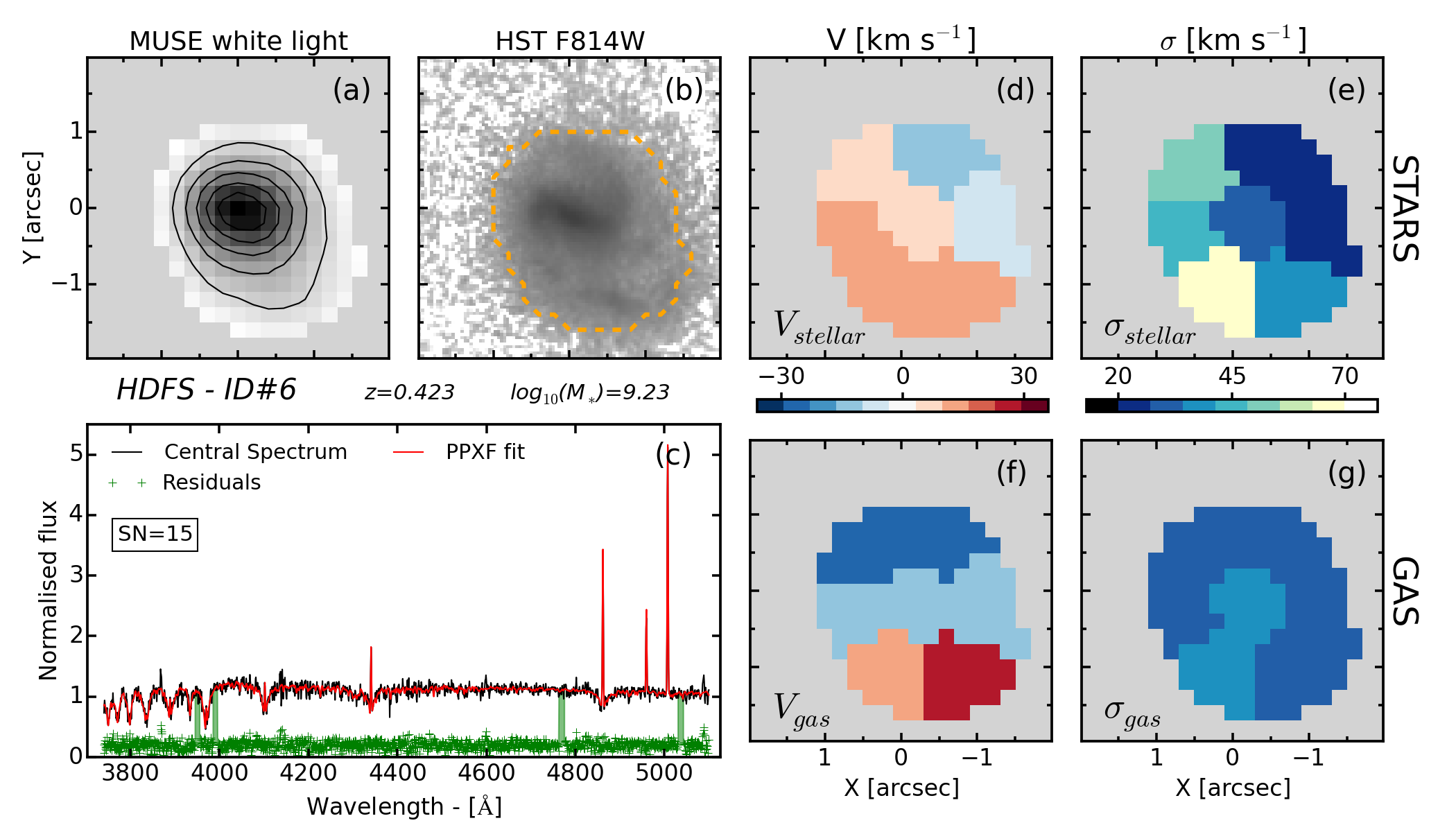}
        \caption{\id6 - caption as Fig.~\ref{fig:Main_kin_fig} }
        \label{fig:kin_maps6}
\end{figure*}

\begin{figure*}
        \includegraphics[width=\textwidth]{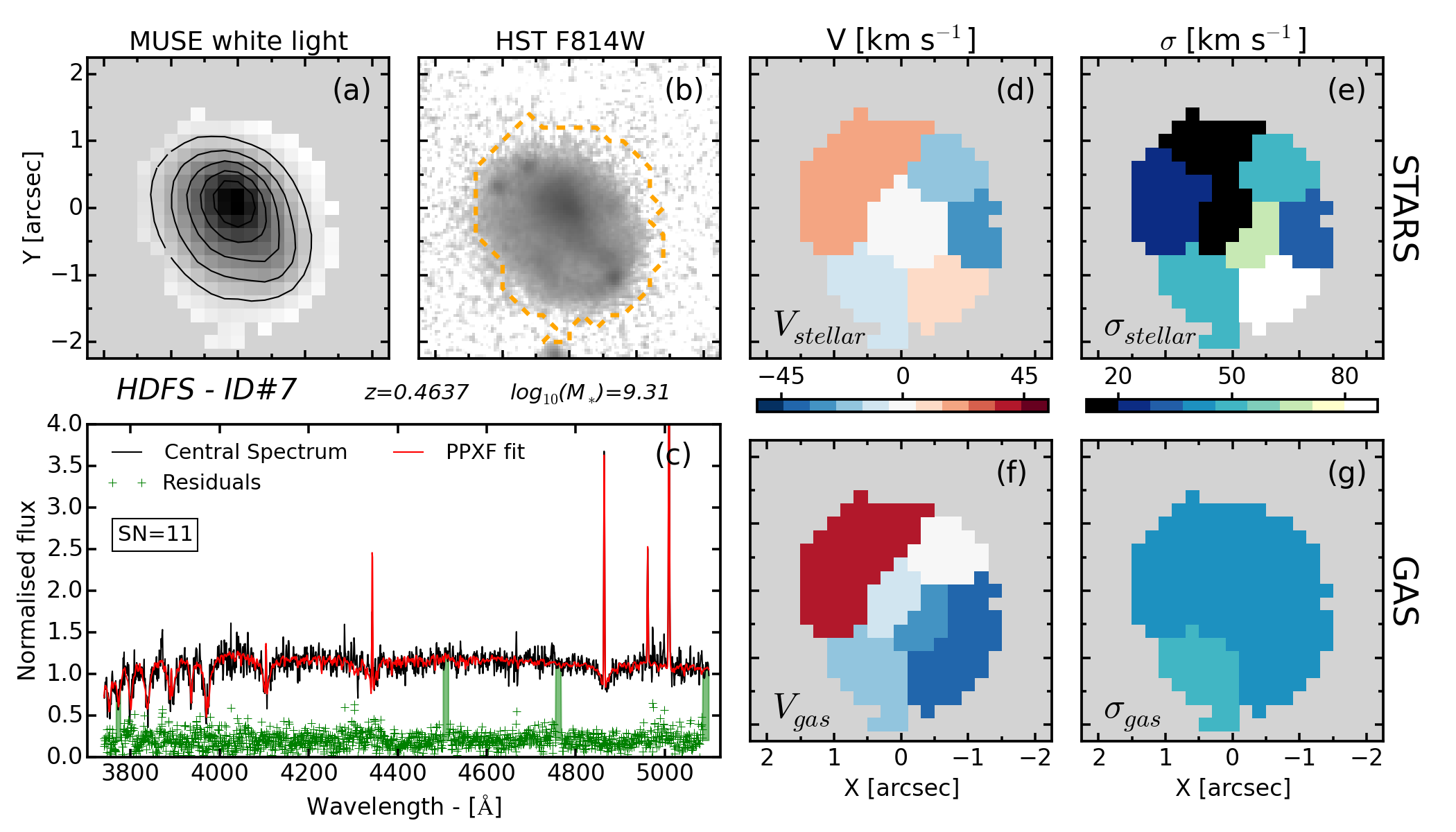}
        \caption{\id7 - caption as Fig.~\ref{fig:Main_kin_fig}}
        \label{fig:kin_maps7}
\end{figure*}

\begin{figure*}
        \includegraphics[width=\textwidth]{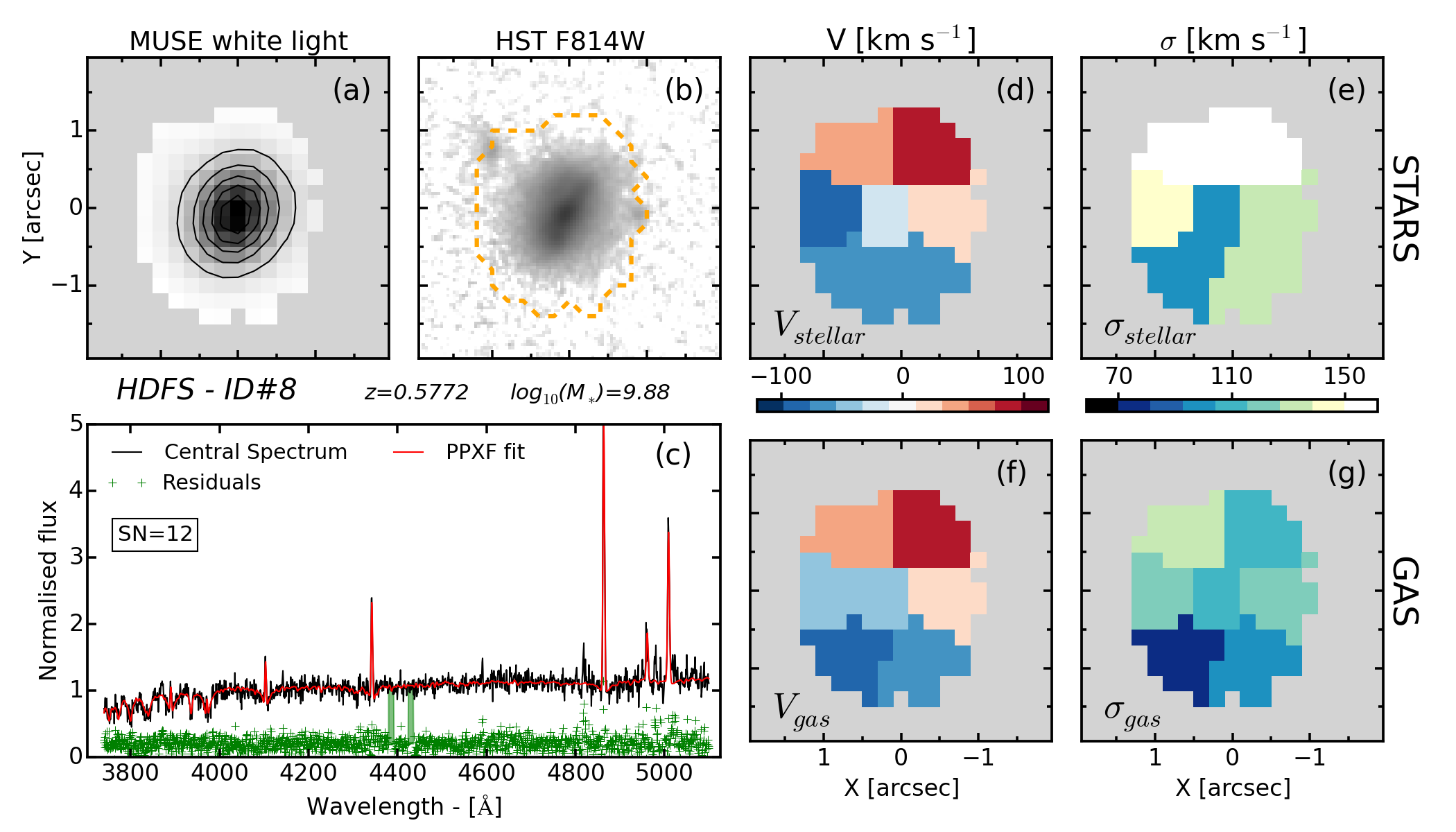}
        \caption{\id8 - caption as Fig.~\ref{fig:Main_kin_fig}}
        \label{fig:kin_maps8}
\end{figure*}

\begin{figure*}
        \includegraphics[width=\textwidth]{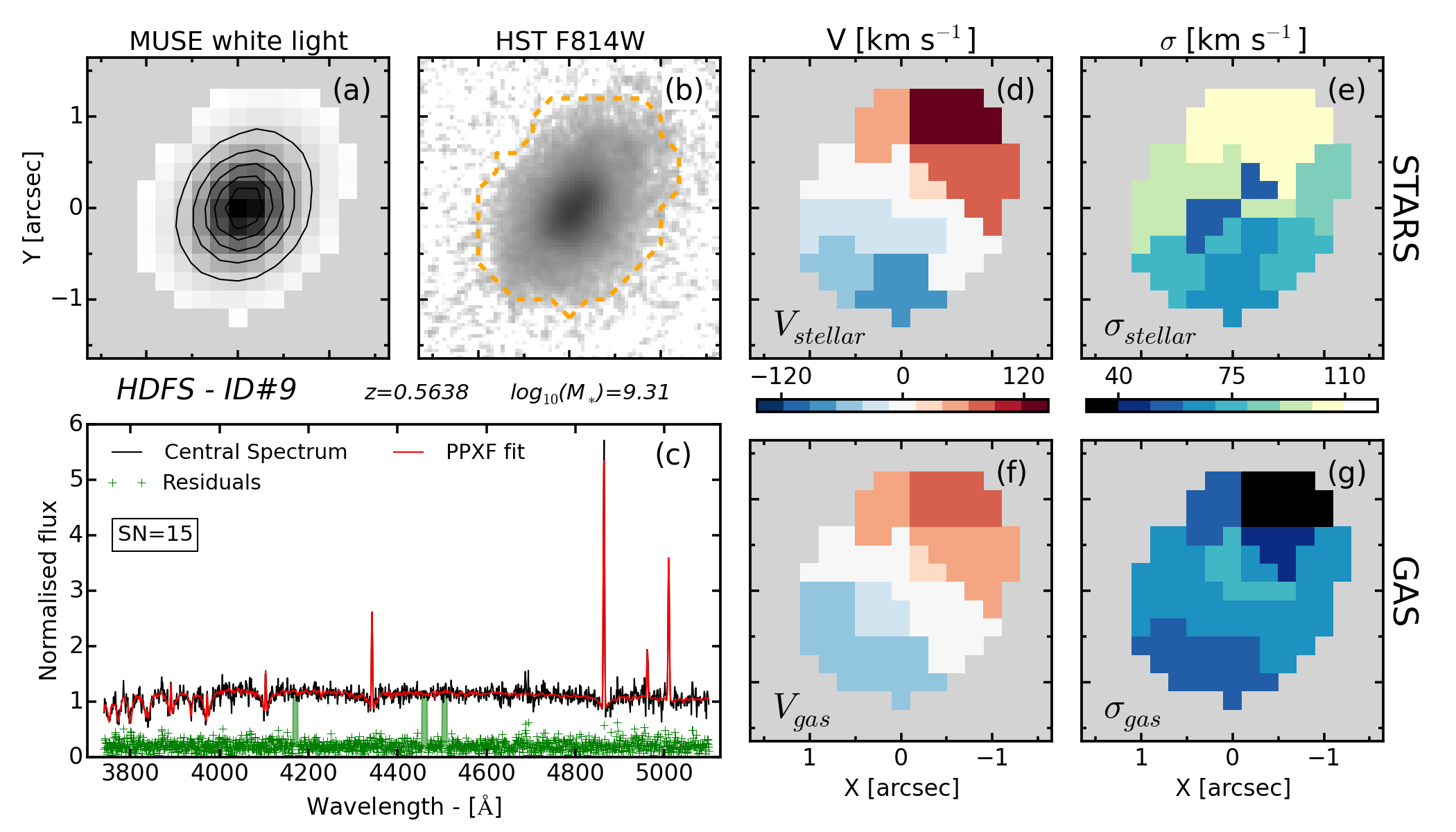}
        \caption{\id9 - caption as Fig.~\ref{fig:Main_kin_fig}}
        \label{fig:kin_maps9}
\end{figure*}

\begin{figure*}
        \includegraphics[width=\textwidth]{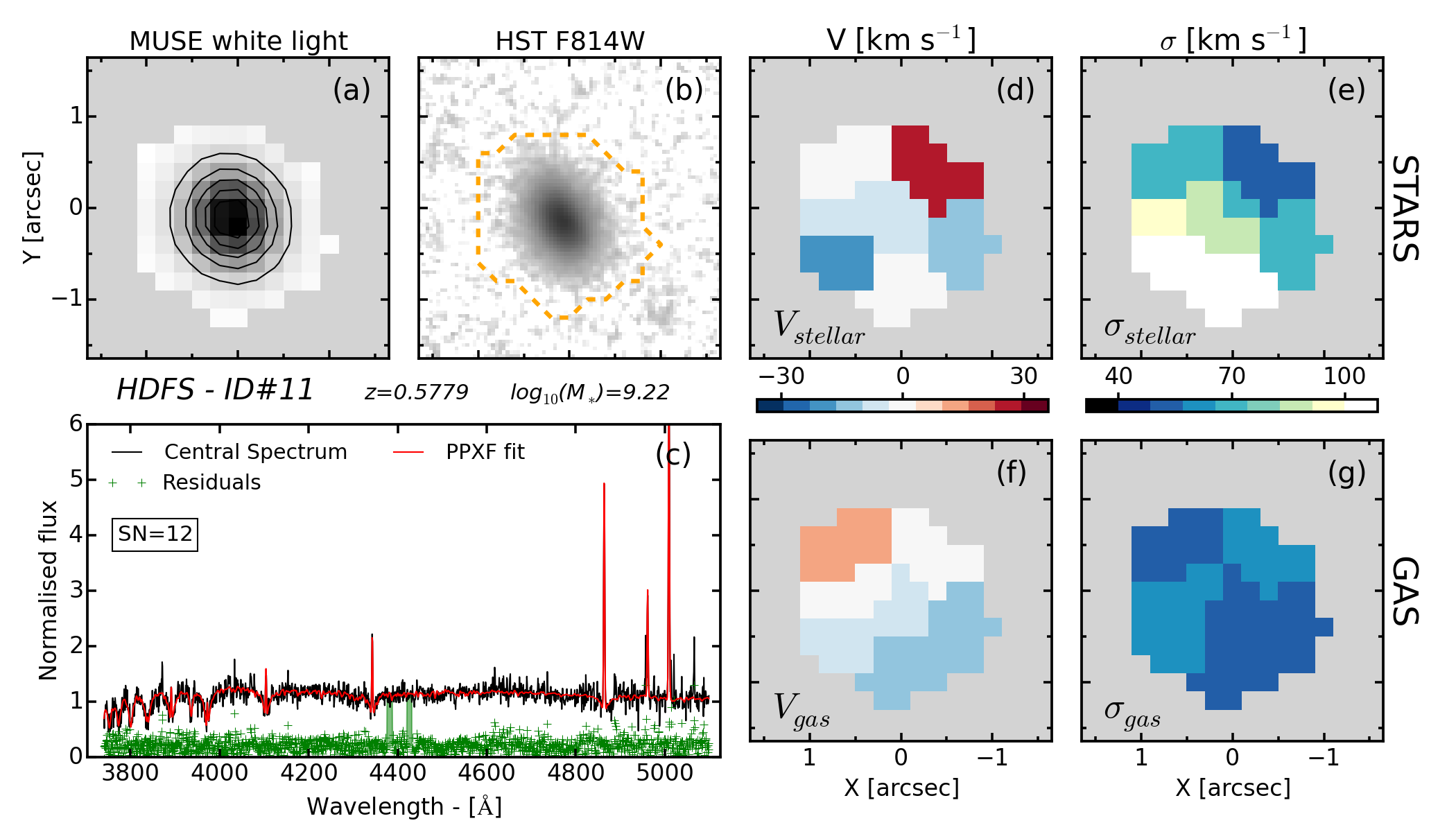}
        \caption{\id11 - caption as Fig.~\ref{fig:Main_kin_fig}}
        \label{fig:kin_maps11}
\end{figure*}

\begin{figure*}
        \includegraphics[width=\textwidth]{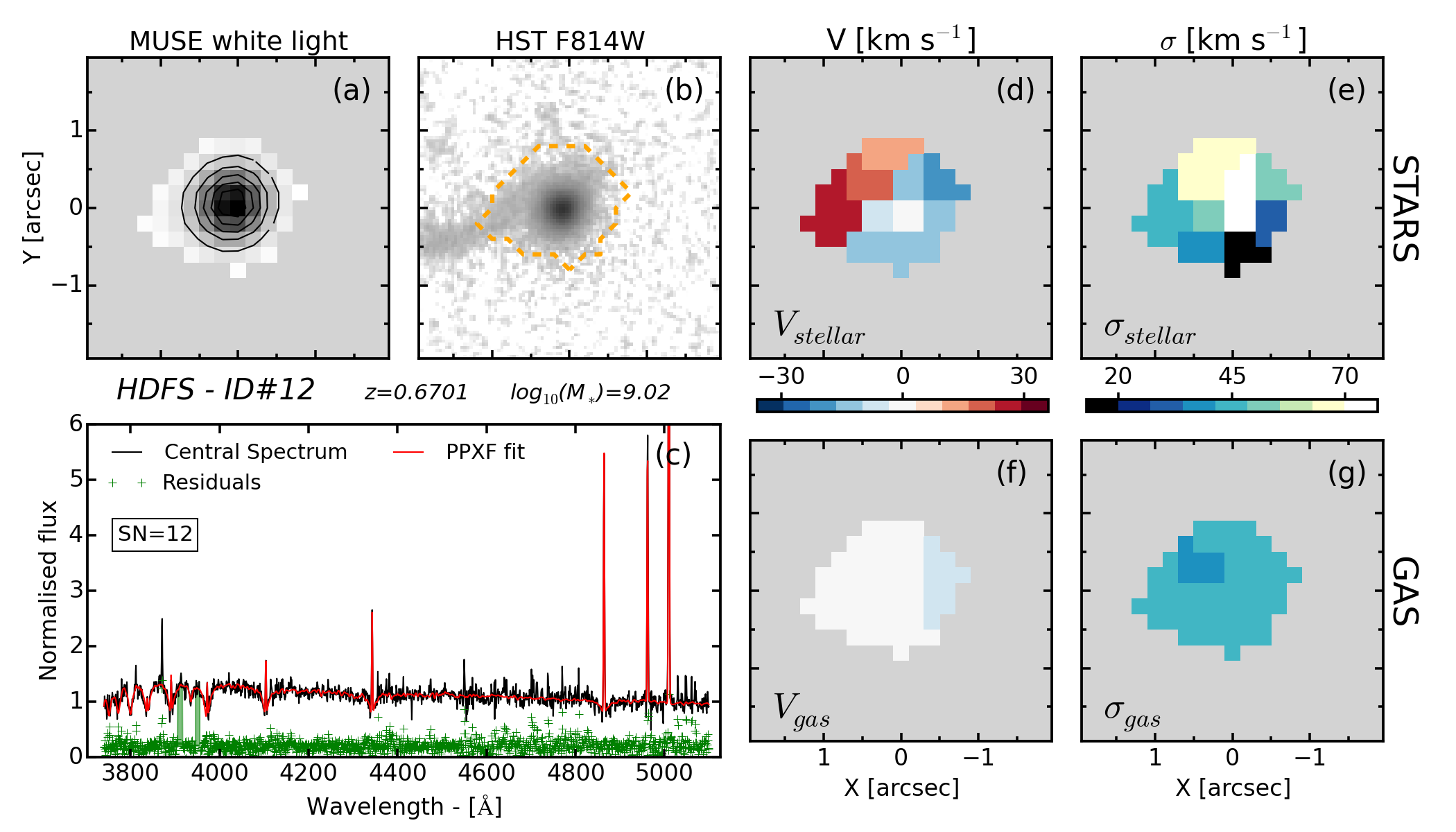}
        \caption{\id12 - caption as Fig.~\ref{fig:Main_kin_fig}}
        \label{fig:kin_maps12}
\end{figure*}

\begin{figure*}
        \includegraphics[width=\textwidth]{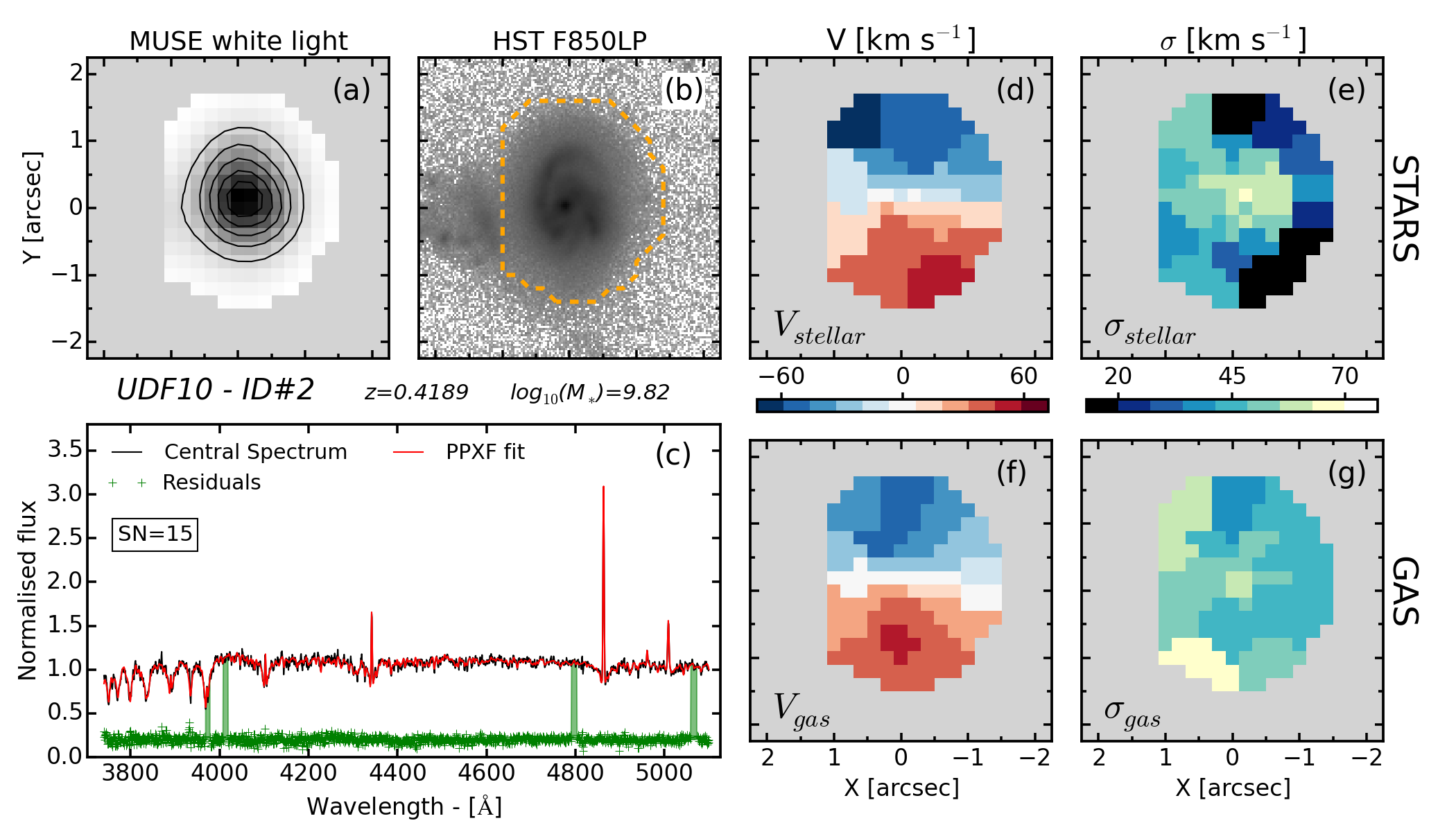}
        \caption{\idu2 - caption as Fig.~\ref{fig:Main_kin_fig}}
        \label{fig:kin_maps2_udf}
\end{figure*}

\begin{figure*}
        \includegraphics[width=\textwidth]{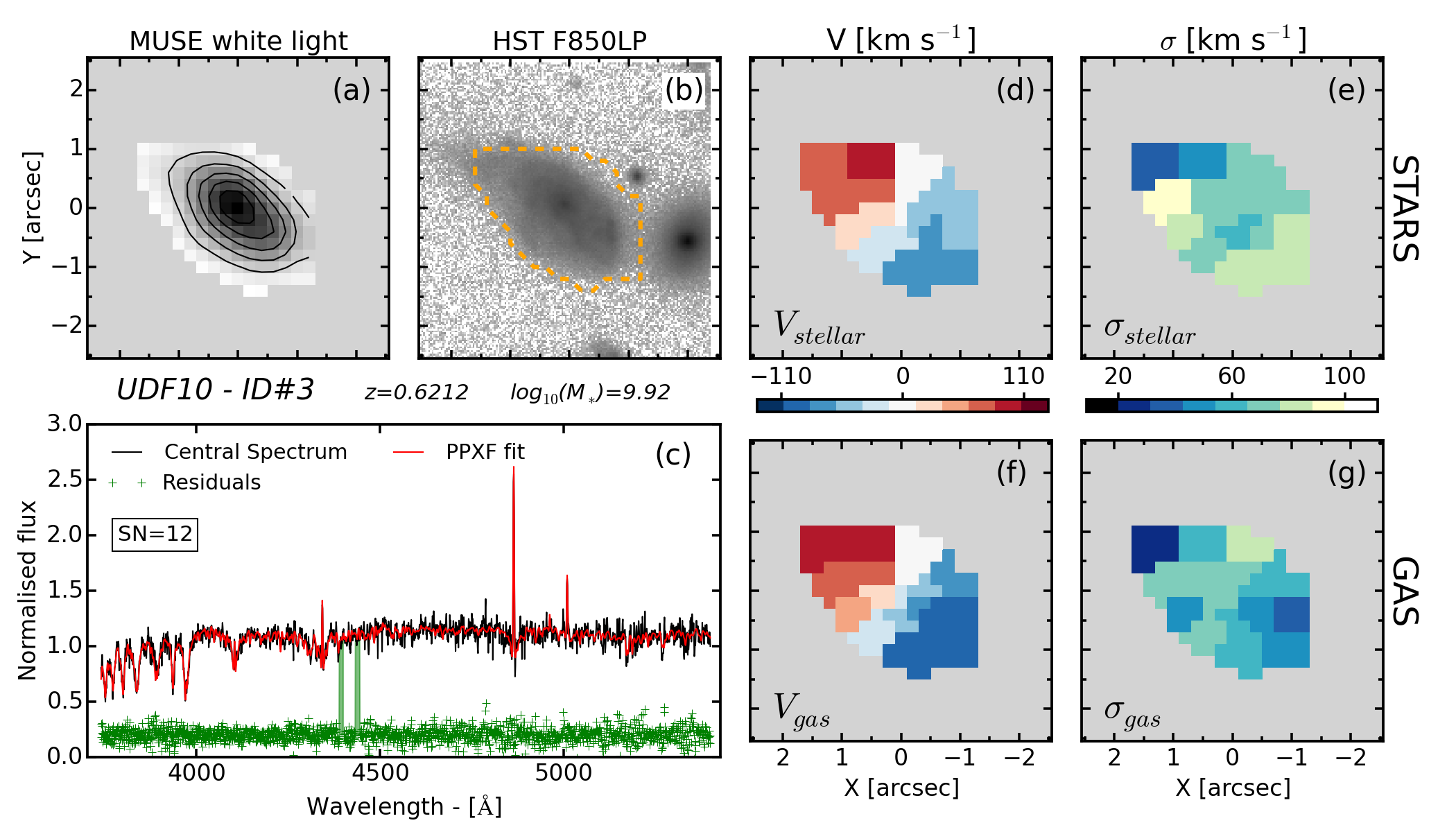}
        \caption{\idu3 - caption as Fig.~\ref{fig:Main_kin_fig}}
        \label{fig:kin_maps3_udf}
\end{figure*}

\begin{figure*}
        \includegraphics[width=\textwidth]{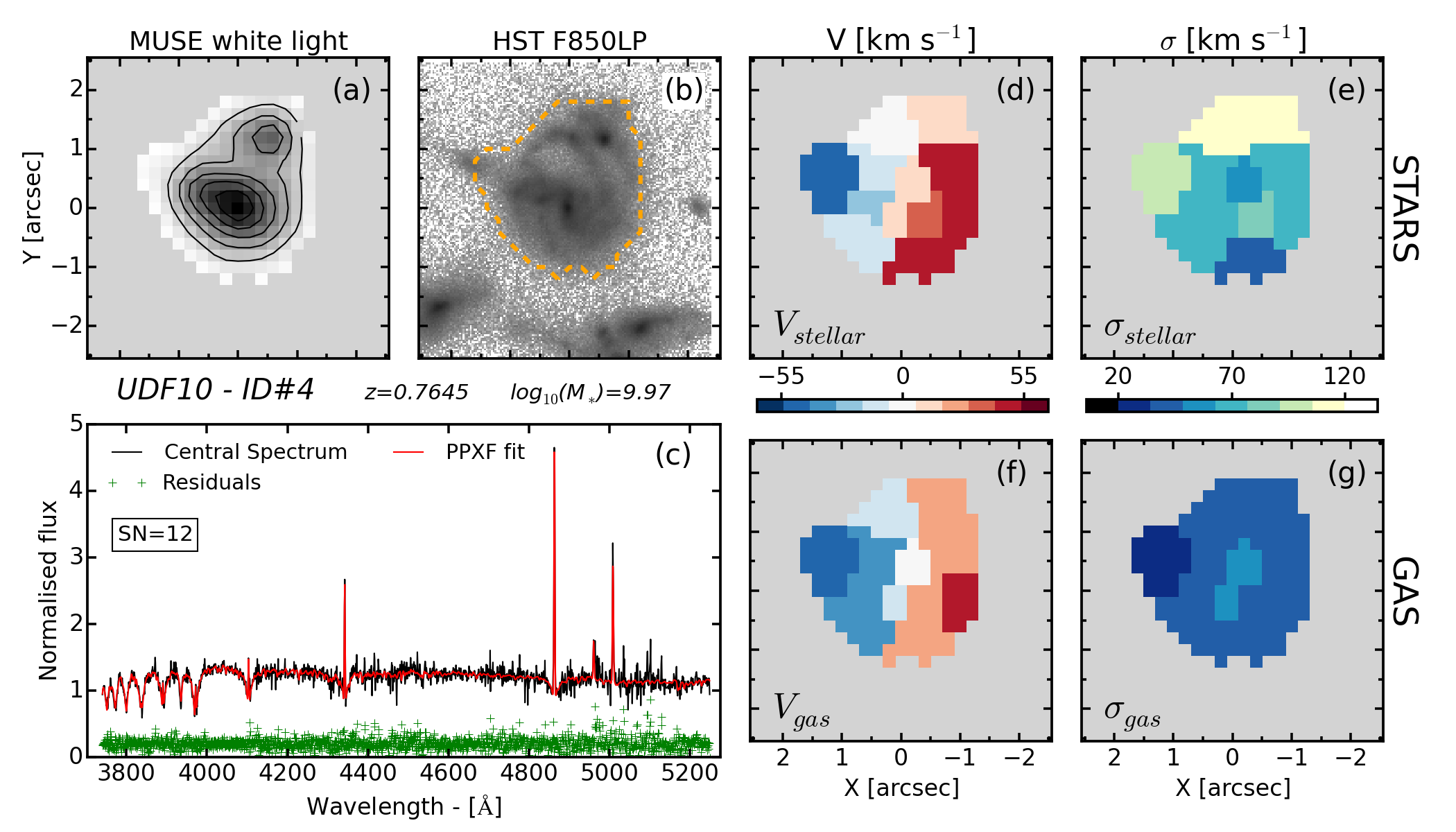}
        \caption{\idu4 - caption as Fig.~\ref{fig:Main_kin_fig}}
        \label{fig:kin_maps4_udf}
\end{figure*}

\begin{figure*}
        \includegraphics[width=\textwidth]{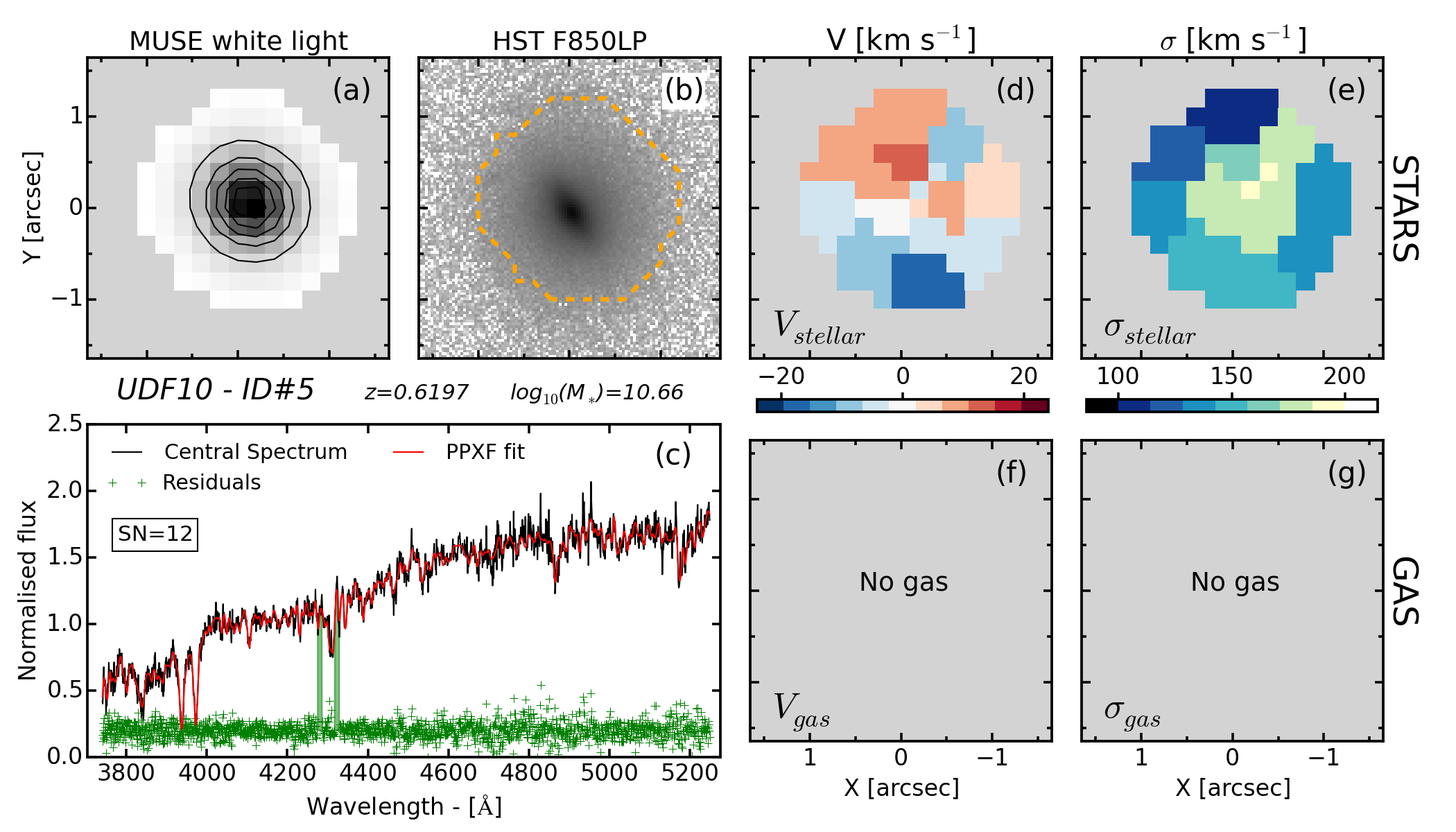}
        \caption{\idu5 - caption as Fig.~\ref{fig:Main_kin_fig}}
        \label{fig:kin_maps5_udf}
\end{figure*}

\begin{figure*}
        \includegraphics[width=\textwidth]{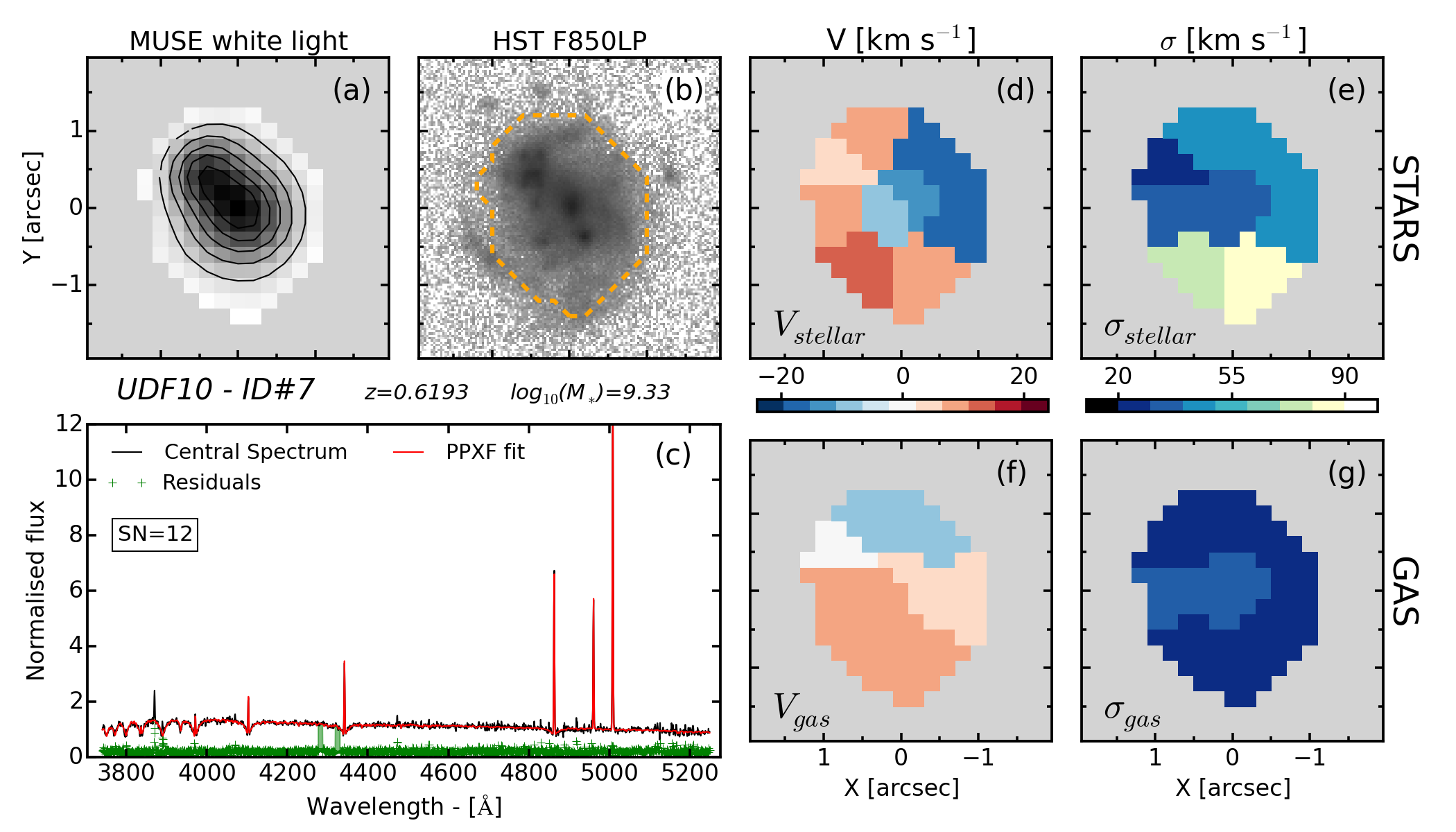}
        \caption{\idu7 - caption as Fig.~\ref{fig:Main_kin_fig}}
        \label{fig:kin_maps7_udf}
\end{figure*}

\begin{figure*}
        \includegraphics[width=\textwidth]{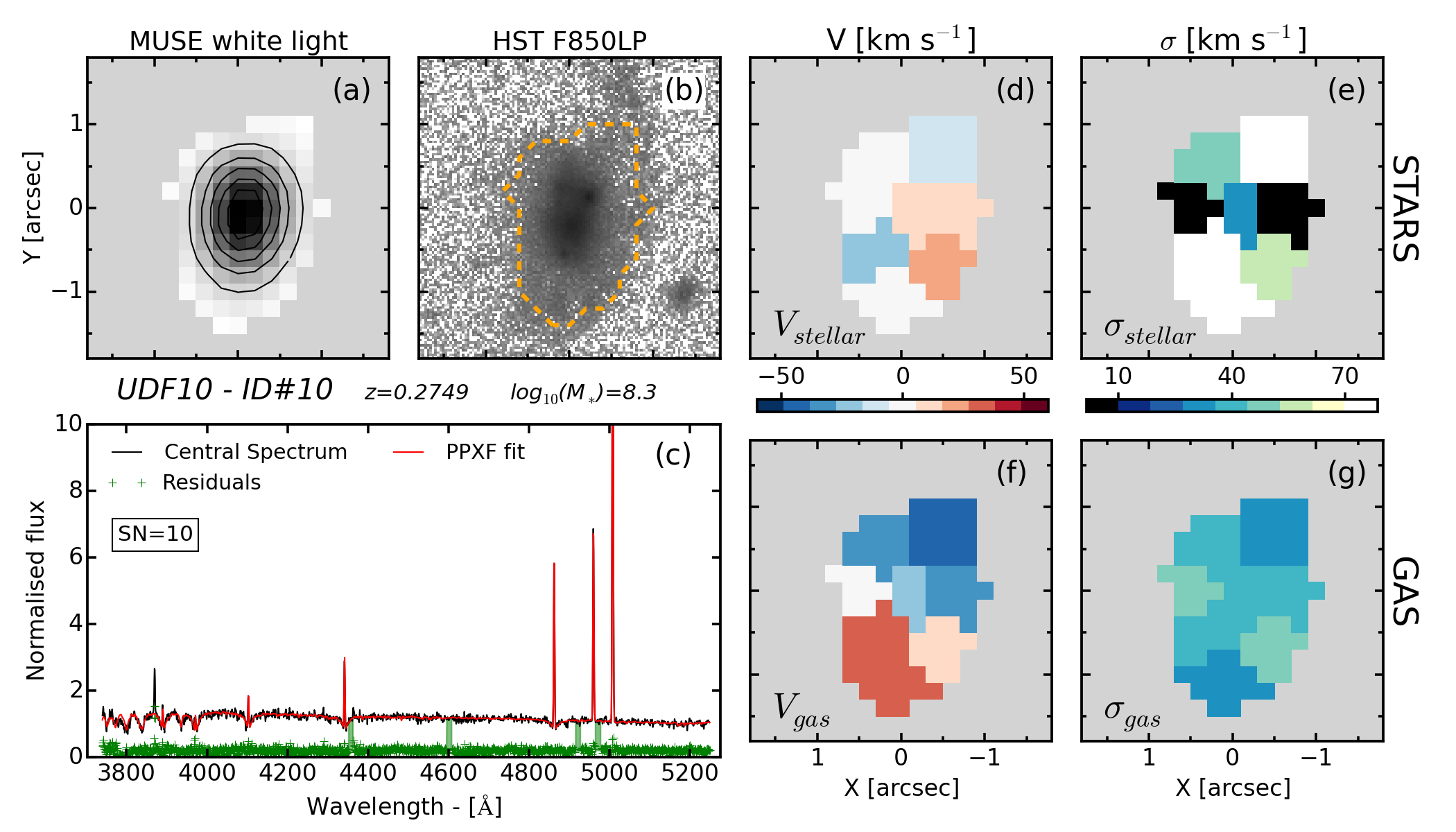}
        \caption{\idu10 - caption as Fig.~\ref{fig:Main_kin_fig}. Despite [OIII] being the strongest line, we fitted the Balmers series due to strong sky line contamination of the [OIII] lines of the outer spatial bins.} 
        \label{fig:kin_maps10_udf}
\end{figure*}

\FloatBarrier
\section{One-dimensional rotation curves \& velocity dispersion profiles}
\label{sec:appendixB}

\begin{figure}
        \includegraphics[width=\columnwidth]{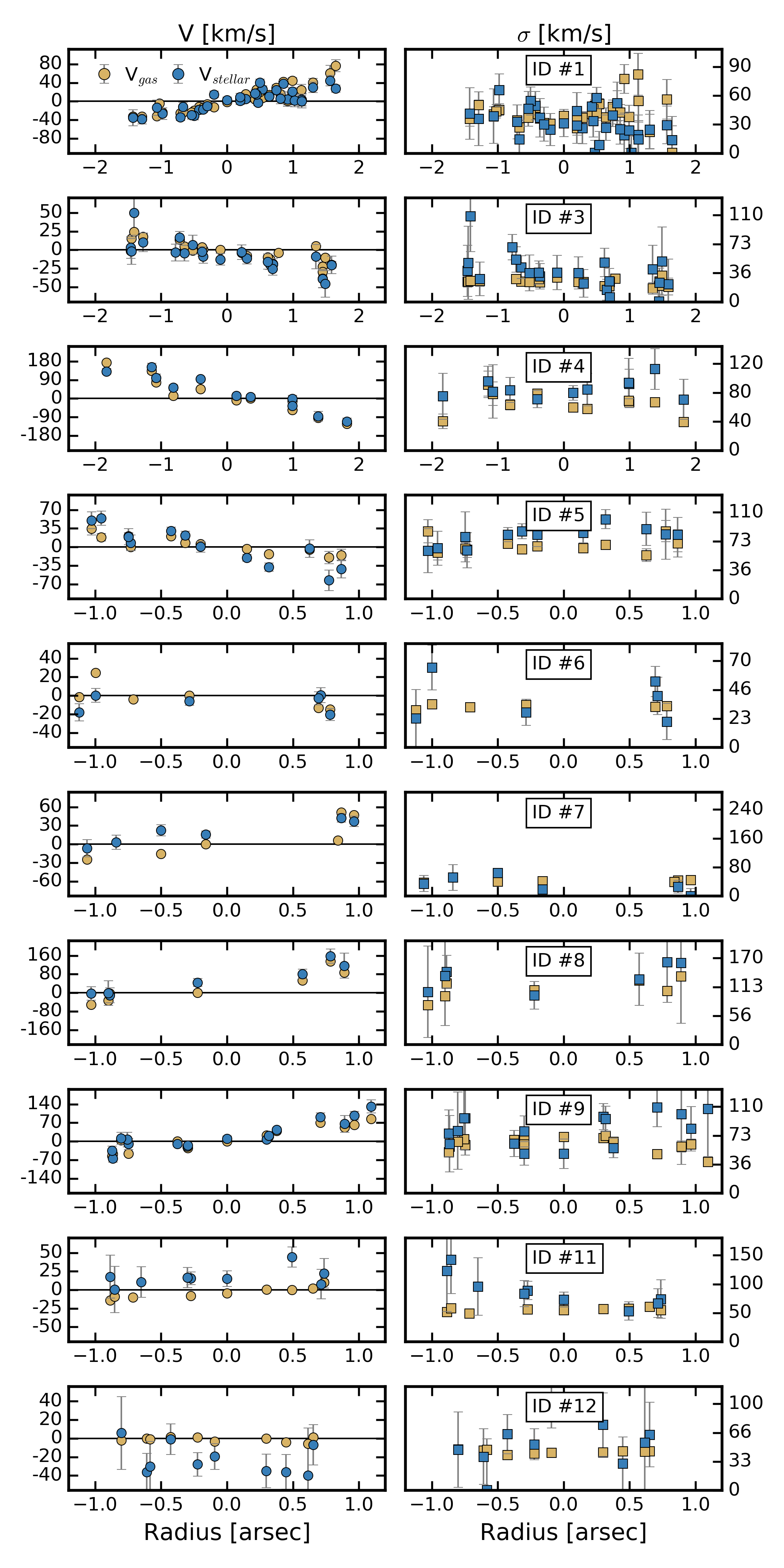}
        \caption{One-dimensional version of the kinematic maps for the HDFS galaxies. Each line corresponds to one galaxy, indicated by the ID number on the right panel. The left column shows the LOS velocity, $V$, and the right column shows the LOS velocity dispersion, $\sigma$. Each point corresponds to a spatial bin of the MUSE sub-cubes and the radius is the distance from the center (i.e.\, not interpolated either cut along the major axis). Blue points are for the stellar component and orange points are for the gas.}
        \label{fig:rot_curve_hdfs}
\end{figure}

\begin{figure}
        \includegraphics[width=\columnwidth]{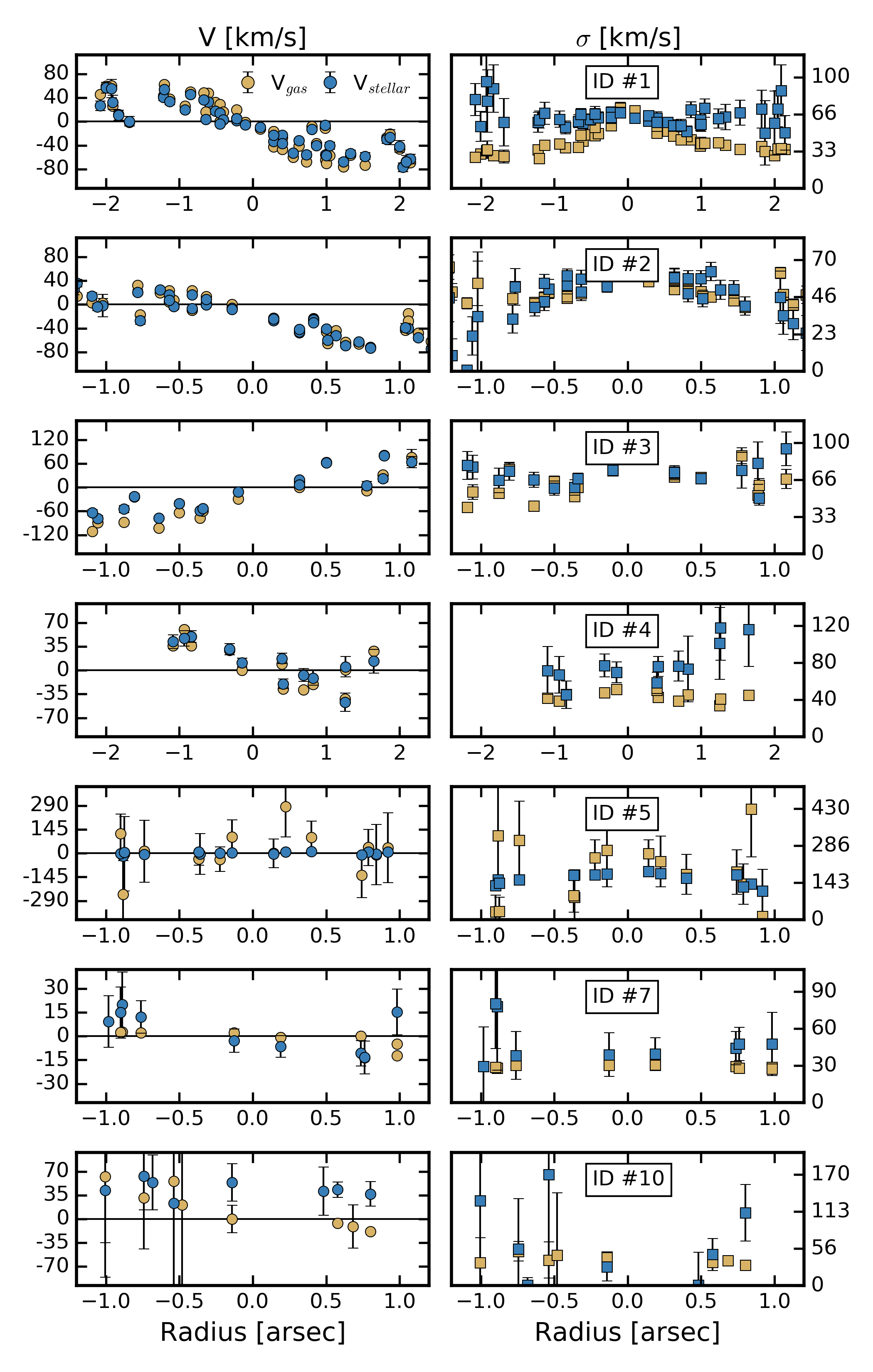}
        \caption{Same as \ref{fig:rot_curve_hdfs} for the \udft~galaxies.}
        \label{fig:rot_curve_udf}
\end{figure}

\end{appendix}

\end{document}